\documentclass[12pt,a4paper]{article}
\pdfoutput=1
\usepackage{jheppub}
\usepackage{amsmath,amssymb,bbm,slashed,cancel}
\usepackage{graphicx}
\usepackage{slashed}
\usepackage{hyperref}
\usepackage{subcaption}
\usepackage{dsfont}
\usepackage{stackengine}
\usepackage{accents}
\usepackage[utf8]{inputenc}
\usepackage{multirow}
\usepackage{physics}

\usepackage{comment}

\usepackage{colortbl,colordvi,xcolor,color}

\usepackage{array,multirow}
\usepackage{booktabs}
\usepackage{float}

\allowdisplaybreaks

\newcommand{\eg}{{\em e.g.}}
\newcommand{\ie}{{\em i.e.}}
\newcommand\unit[1]{\,\mathrm{#1}}
\newcommand\MeV{\unit{MeV}}
\newcommand\GeV{\unit{GeV}}
\newcommand\TeV{\unit{TeV}}

\makeatletter
\gdef\@fpheader{}
\makeatother

\preprint{KEK--TH--2497}

\title{Electroweak precision test of axion-like particles}
\author[a]{Masashi Aiko}
\author[a,b]{and Motoi Endo}

\affiliation[a]{KEK Theory Center, IPNS, KEK, Tsukuba, Ibaraki 305--0801, Japan}
\affiliation[b]{The Graduate University of Advanced Studies (Sokendai), Tsukuba, Ibaraki 305--0801, Japan}

\abstract{
We study the contributions of an axion-like particle to the electroweak precision observables.
The particle is assumed to couple with the standard model electroweak gauge bosons.
We provide the formulae of the contributions valid for any mass of the axion-like particle. 
It is found that the effects arise not only via the oblique $S$ and $U$ parameters but also via radiative corrections to the gauge couplings.
Besides, the decay of $Z\to a\gamma$ affects the total width of the $Z$ boson.
All of those contributions are considered simultaneously in the global fit analysis of the electroweak precision observables.
Also, we discuss the recent CDF result of the $W$-boson mass measurement. 
Since the model is tightly constrained by flavor and collider constraints, it is found that the discrepancy from the standard model prediction is solved only when the axion-like particle is heavier than $500\GeV$ and its coupling to di-photon is suppressed.
}

\begin{document}
\maketitle
\flushbottom

\renewcommand{\thefootnote}{\#\arabic{footnote}}
\setcounter{footnote}{0}

\section{Introduction} \label{sec: intro}

New light pseudoscalar particles are one of the most popular and simple extensions of the Standard Model (SM).
They are often motivated in models with spontaneous violations of global symmetries and are called axion-like particles (ALPs).
Their masses are generated by explicit small violations of the global symmetries, whereas their interactions are characterized by the symmetries. 
They affect low-energy observables via the interactions with the SM particles.

In this paper, we revisit contributions to the electroweak precision observables (EWPOs).
The ALPs are assumed to be coupled primarily with the SM ${\rm U(1)}_Y$ and ${\rm SU(2)}_L$ gauge bosons, yielding ALP couplings with the photon ($\gamma$) and $Z$ boson as well as the charged $W$ boson after the electroweak (EW) symmetry breaking.
Then, the EWPOs are affected via vacuum polarizations of $\gamma$, $Z$, and $W$. 
Such contributions have been studied in Ref.~\cite{Bauer:2017ris}. 
The authors argued that they are expressed by the oblique parameters, $S,~T$, and $U$~\cite{Peskin:1991sw}; $S$ and $U$ are generated, while $T$ is absent at least at the one-loop level.
Interestingly, $U$ becomes comparable to $S$, unlikely to a wide class of new physics models. 
Also, the CDF~II collaboration recently reported a new result of the $W$ mass measurement~\cite{CDF:2022hxs}. 
The result is not consistent with the SM prediction as well as the previous experimental values. 
In Ref.~\cite{Yuan:2022cpw}, the CDF result has been discussed in the ALP models based on Ref.~\cite{Bauer:2017ris}, and it was concluded that a light ALP can solve the disagreement marginally.

It is noticed that the above studies ignored the other ALP contributions, \ie, those except for $S,~T$, and $U$. 
In many analyses of the oblique parameters, \eg, those in Refs.~\cite{Baak:2014ora, ParticleDataGroup:2022pth}, $S,~T$, and $U$ have been restricted/determined by fitting them globally to the EWPOs under the assumption that there are no additional contributions from new physics.
However, we will show in this paper that this assumption is not valid in the ALP models.
These three parameters are not enough to parameterize the ALP effects, but there are additional contributions via vacuum polarizations such as the oblique parameters beyond $S$, $T$, and $U$ (cf., Refs.~\cite{Maksymyk:1993zm, Barbieri:2004qk}). 
Although these extra contributions are suppressed in a wide class of models, this is not the case for the ALP; they can be comparable to $S$ and $U$. 
Furthermore, the $Z$ boson can decay into a light ALP and a photon.
This decay proceeds at the tree level and contributes to the total width of the $Z$ boson. 
Since those contributions affect the EWPOs simultaneously with $S$, $T$, and $U$, they must be analyzed {\it collectively} in the global fit.
It will be shown that the global-fit results are changed drastically. 

The ALPs are subject to experimental constraints. 
Although cosmological limits are very severe, they can be avoided if the ALPs are heavier than $\sim 1\GeV$~\cite{Jaeckel:2010ni, Cadamuro:2011fd, Proceedings:2012ulb}. 
Even in such a case, the ALPs affect meson decays via the interactions with the $W$ boson~\cite{Izaguirre:2016dfi, Alonso-Alvarez:2018irt, Gavela:2019wzg, Guerrera:2021yss, Bauer:2021mvw, Guerrera:2022ykl}. 
In particular, the $B$-meson decay into a $K^{(*)}$ meson with photons via $a \to \gamma\gamma$ or the decay with leptons via $a \to \ell^+\ell^-$ is very sensitive to the ALP contributions. 
Besides, the ALPs have been studied particularly in the LEP and LHC experiments~\cite{Mimasu:2014nea, Jaeckel:2015jla, Jaeckel:2012yz, Bauer:2017ris, Bauer:2018uxu, Florez:2021zoo, Wang:2021uyb, dEnterria:2021ljz, Knapen:2016moh, CMS:2018erd, ATLAS:2020hii, Craig:2018kne, Bonilla:2022pxu}. 
In this paper, their results are applied to the current model setup, and we will compare all those constraints with the EWPO fit results. 
The recent CDF result of the $W$ mass measurement~\cite{CDF:2022hxs} will also be discussed under those constraints.

This paper is organized as follows.
In Sec.~\ref{sec: alp}, we introduce the ALP model and provide its decay rates.
In Sec.~\ref{sec: ewpt}, we explain the ALP contributions to the EWPOs and the analysis strategy.
The experimental constraints are summarized in Sec.~\ref{sec: constraint}.
We show the numerical results in Sec.~\ref{sec: result}, and Sec.~\ref{sec: conclusion} is devoted to the conclusion. 
In Appendix~\ref{app: PV_funcs}, the Passarino-Veltman function~\cite{Passarino:1978jh} is given explicitly.
In Appendix~\ref{app: aToZstarA}, we give the analytic formula of the three-body decay width for $a\to Z^{*}\gamma$. 

\section{Model} \label{sec: alp}

We consider an ALP $(a)$ coupled with the ${\rm SU(2)}_L$ gauge boson $(W_\mu^a)$ and the ${\rm U(1)}_Y$ gauge boson $(B_\mu)$.
The Lagrangian is shown as~\cite{Georgi:1986df}
\begin{align}
\mathcal{L}_{\mathrm{ALP}}
=
\frac{1}{2}\partial_{\mu}a\partial^{\mu}a
-\frac{1}{2} m_{a}^{2} a^{2}
-c_{WW}\frac{a}{f_{a}}W_{\mu\nu}^{a}\widetilde{W}^{a\mu\nu}
-c_{BB}\frac{a}{f_{a}}B_{\mu\nu}\widetilde{B}^{\mu\nu},
\label{eq: Lagrangian}
\end{align}
where $m_a$ is the ALP mass, and $f_a$ is the ALP decay constant.
The coefficients, $c_{WW}$ and $c_{BB}$, as well as $m_a$ and $f_a$ are regarded as free parameters, though $m_a < f_a$ is satisfied.  
We set $f_a = 1\TeV$ throughout this paper.
The field strengths of the ${\rm SU(2)}_{L}$ and ${\rm U(1)}_{Y}$ gauge bosons are defined as
\begin{align}
W_{\mu\nu}^{a} &= \partial_{\mu}W_{\nu}^{a}-\partial_{\nu}W_{\mu}^{a}+g\epsilon^{abc}W_{\mu}^{b}W_{\nu}^{c}, \\
B_{\mu\nu} &= \partial_{\mu}B_{\nu}-\partial_{\nu}B_{\mu},
\end{align}
where $g$ is the ${\rm SU(2)}_L$ gauge coupling constant. 
The dual is expressed by 
\begin{align}
\widetilde{X}_{\mu\nu} = \frac{1}{2}\epsilon^{\mu\nu\rho\sigma}X_{\rho\sigma}\qc
(X = W^{a}, B).
\end{align}
The totally antisymmetric tensors are defined with $\epsilon^{012}=1$ and $\epsilon^{0123}=1$.
After the EW symmetry breaking, the above interactions are rewritten as
\begin{align}
\mathcal{L}_{\mathrm{int}}
&=
-\frac{1}{4}g_{a\gamma\gamma}aF_{\mu\nu}\widetilde{F}^{\mu\nu}
-\frac{1}{2}g_{aZ\gamma}aZ_{\mu\nu}\widetilde{F}^{\mu\nu} \notag \\
&\quad
-\frac{1}{4}g_{aZZ}aZ_{\mu\nu}\widetilde{Z}^{\mu\nu}
-\frac{1}{2}g_{aWW}aW_{\mu\nu}^{+}\widetilde{W}^{-\mu\nu}
+ \ldots,
\label{eq: effective_coupling}
\end{align}
where quartic interaction terms are omitted. 
Here, the field strengths are given by
\begin{align}
X'_{\mu\nu} = \partial_{\mu}X'_{\nu}-\partial_{\nu}X'_{\mu}.
\end{align}
where $X'_{\mu} = A_\mu$ for the photon $(\gamma)$, $Z_\mu$ for the $Z$ boson, and $W_\mu^{\pm}$ for the charged $W$ boson. 
Note that $F_{\mu\nu}$ corresponds to $A_\mu$.
Then, the coupling constants are expressed by $c_{WW}$ and $c_{BB}$ with $f_a$ as
\begin{align}
g_{a\gamma\gamma} &= \frac{4}{f_{a}}\qty(s_{W}^{2}c_{WW}+c_{W}^{2}c_{BB}), 
\label{eq: gaAA} \\
g_{aZ\gamma} &= \frac{2}{f_{a}}\qty(c_{WW}-c_{BB})s_{2W}, 
\label{eq: gaZA} \\
g_{aZZ} &= \frac{4}{f_{a}}\qty(c_{W}^{2}c_{WW}+s_{W}^{2}c_{BB}), 
\label{eq: gaZZ} \\
g_{aWW} &= \frac{4}{f_{a}}c_{WW},
\label{eq: gaWW}
\end{align}
where $c_{W} = \cos{\theta_{W}}$, $s_{W} = \sin{\theta_{W}}$, and $s_{2W} = \sin{2\theta_{W}}$ with the Weinberg angle $\theta_{W}$.

\subsection{Decay of ALP} \label{sec: decay}

The ALP decays into a pair of SM gauge bosons.
The partial decay widths for $a\to V_{i}V_{j}\ (V_{i,j}=\gamma,Z,W^{\pm})$ are obtained as~\cite{Bauer:2017ris, Craig:2018kne, Bonilla:2021ufe}
\begin{align}
    \Gamma(a\to V_{i}V_{j}) = \frac{m_{a}^{3}}{32\pi(1+\delta_{ij})}\lambda^{3/2}
    \qty(\frac{m_{V_{i}}^{2}}{m_{a}^{2}}, \frac{m_{V_{j}}^{2}}{m_{a}^{2}})\abs{g_{aV_{i}V_{j}}^{\rm eff}}^{2},
    \label{eq: decay_a_VV}
\end{align}
with $\lambda(x,y) = (1-x-y)^{2}-4xy$.
Here $m_{V}$ is the gauge-boson mass ($m_\gamma=0$).
Note that $\delta_{ij}=0$ for $a\to Z\gamma$ and $a\to W^+ W^-$.

The ALP coupling to the SM gauge bosons $g^{\rm eff}_{aV_{i}V_{j}}$ is obtained both at the tree and loop levels.
In particular, the coupling to the di-photon is given at the tree level by $g_{a\gamma\gamma}$ in Eq.~\eqref{eq: gaAA} and is generated by loop corrections with $g_{aWW}$ in Eq.~\eqref{eq: gaWW} as~\cite{Bauer:2017ris}
\begin{align}
    g_{a\gamma\gamma}^{\rm eff} = g_{a\gamma\gamma}+\frac{2\alpha}{\pi} g_{aWW}B_{2}(\tau_{W}),
\end{align}
where $\alpha\equiv e^{2}/(4\pi)$ with the QED coupling $e=g s_{W}$.
The loop function $B_{2}$ is defined as
\begin{align}
    B_{2}(\tau) = 1-(\tau-1)f^{2}(\tau),
\end{align}
where $f(\tau)$ is given by
\begin{align}
    f(\tau) =
    \left\{\mqty{
        \arcsin{\frac{1}{\sqrt{\tau}}}, \\
        \frac{\pi}{2}+\frac{i}{2}\log{\frac{1+\sqrt{1-\tau}}{1-\sqrt{1-\tau}}},
    }\right.\qquad
    \mqty{
        \text{for}~\tau\geq 1\\
        \text{for}~\tau< 1
    },
\end{align}
with $\tau_{W}=4m_{W}^{2}/m_{a}^{2}$.
In the light ALP limit, $\tau_{W}\gg 1$, the function is approximated as $B_{2}\to m_{a}^{2}/(6m_{W}^{2})$, \ie, suppressed by $m_{W}$.
In the heavy limit, $\tau_{W}\ll 1$, we obtain $B_{2}\to 1+\pi^{2}/4-\log^{2}{(m_{a}/m_{W})}$, which is enlarged by the logarithm.
On the other hand, it is sufficient for us to evaluate the decay widths for $a\to Z\gamma,\, ZZ,\, W^{+}W^{-}$ at the tree level, \ie, $g_{aV_{i}V_{j}}^{\rm eff} = g_{aV_{i}V_{j}}$.

When the ALP is lighter than the $Z$ boson, and if $g_{a\gamma\gamma}^{\rm eff}$ is suppressed, the ALP decays into either a pair of SM fermions, $a \to f\bar f$, or the fermions with a photon, $a \to Z^{*}\gamma\to f\bar f \gamma$, by exchanging an off-shell $Z$ boson. 
Since the ALP does not couple with SM fermions directly, the former proceeds via radiative corrections, as will be explained later.
On the other hand, even though the latter is a three-body decay, it proceeds at the tree level, and thus, its decay width can be comparable to the former. 
For $m_a \ll m_{Z}$, the decay width is obtained as
\begin{align}
 \Gamma(a \to f\bar f \gamma) =
 N_{c}^{f}\frac{g_{Z}^{2}g_{aZ\gamma}^{2}}{30720\pi^{3}}
 \qty[(g_{V,f})^{2}+(g_{A,f})^{2}]\frac{m_{a}^{7}}{m_{Z}^{4}},
\end{align}
where $N_{c}^{f} = 1~ (3)$ is the color factor for leptons (quarks), and $g_Z = g/c_{W}$ is the $Z$-boson coupling constant.
The vector and axial form-factors of $Z$ boson are defined as $g_{V,f} = I_{3}^f - 2\, Q_f s_W^2$ and $g_{A,f} = I_{3}^f$.
The formula valid for any $m_a$ is provided in Appendix~\ref{app: aToZstarA}.

The ALP decays into a pair of leptons and heavy quarks, $a\to f\bar{f}$, via gauge-boson loops.
The partial decay widths are obtained as~\cite{Bauer:2017ris, Bauer:2020jbp}
\begin{align}
    \Gamma(a\to f\bar{f}) =
    N_{c}^{f}\frac{m_{a}m_{f}^{2}}{8\pi}
    \abs{g_{aff}^{\rm eff}}^{2}
    \sqrt{1-\frac{4m_{f}^{2}}{m_{a}^{2}}}.
\end{align}
The effective coupling induced by radiative corrections is given by
\begin{align}
    g_{aff}^{\rm eff} &=
    3Q_{f}^{2}\frac{\alpha}{4\pi}g_{a\gamma\gamma} \ln{\frac{\Lambda^{2}}{m_{f}^{2}}}
    +\frac{3}{4s_{W}^{2}}\frac{\alpha}{4\pi}g_{aWW} \ln{\frac{\Lambda^{2}}{m_{W}^{2}}}
    \notag \\ &\quad
    +\frac{3}{s_{W}c_{W}}\frac{\alpha}{4\pi}g_{aZ\gamma}Q_{f}
        \qty(I_{3}^{f}-2Q_{f}s_{W}^{2})
        \ln{\frac{\Lambda^{2}}{m_{Z}^{2}}}
    \notag \\ &\quad
    +\frac{3}{s_{W}^{2}c_{W}^{2}}\frac{\alpha}{4\pi}g_{aZZ}
        \qty(Q_{f}^{2}s_{W}^{4}-I_{3}^{f}Q_{f}s_{W}^{2}+\frac{1}{8})
        \ln{\frac{\Lambda^{2}}{m_{Z}^{2}}},
    \label{eq: cff_eff}
\end{align}
where $Q_{f}$ and $I_{3}^{f}$ are the electric charge and the weak isospin of the fermion $f$, and $\tau_{f}=4m_{f}^{2}/m_{a}^{2}$.
Here, we keep the terms enhanced by $\ln{(\Lambda^2/m_f^2)}$ or $\ln{(\Lambda^2/m_{V}^2)}$ $(V=Z, W)$.
The cutoff scale $\Lambda$ is determined by a UV theory of the ALP model and treated as a model parameter in this paper.
The subleading terms are given in Refs.~\cite{Bauer:2017ris, Bauer:2020jbp, Bonilla:2021ufe}, though they are irrelevant (cf., Ref.~\cite{Arias-Aragon:2022iwl}).
Note that the above formula is valid for the decay which conserves the lepton/quark flavors. 

The $W$ boson contribution can generate flavor-violating interactions of quarks. 
In particular, those for the down-type quarks can be sizable due to top-quark loops.
With keeping the up-type quark masses non-zero, the following term arises apart from Eq.~\eqref{eq: cff_eff}~\cite{Izaguirre:2016dfi, Alonso-Alvarez:2018irt, Gavela:2019wzg, Guerrera:2021yss, Bauer:2021mvw, Guerrera:2022ykl}:
\begin{align}
    g_{ad_{i}d_{j}}^{\rm eff} = -
    \frac{3}{4s_{W}^{2}}\frac{\alpha}{4\pi}g_{aWW}\sum_{q=u,c,t}V_{qi}V_{qj}^{*}G(x_{q}),
    \label{eq: flavor_violating _coupling}
\end{align}
where $V_{ij}$ is the Cabbibo-Kobayashi-Maskawa (CKM) matrix, and the loop function is defined as
\begin{align}
    G(x) = \frac{x\qty(1-x+x\ln{x})}{(1-x)^{2}}.
\end{align}
with $x_q = m_q^2/m_{W}^2$.

On the other hand, the inclusive rate for the ALP decaying into light hadrons is shown as (cf., Ref.~\cite{Bauer:2017ris})
\begin{align}
    \Gamma(a\to{\rm hadrons}) &=
    32\pi\alpha_{s}^{2}m_{a}^{3}\qty(1+\frac{83}{4}\frac{\alpha_{s}}{\pi})\abs{g_{agg}^{\rm eff}}^{2}
    +\sum_{f=u,d,s}\Gamma(a\to f\bar{f}),
\end{align}
where $m_{a}\gg\Lambda_{\rm QCD}$ is assumed. 
The effective coupling with gluons is given by
\begin{align}
    g_{agg}^{\rm eff}=\frac{1}{32\pi^{2}}\sum_{f=u,d,s}g_{aff}^{\rm eff}.
\end{align}
Note that the above expression is not valid for $m_{a} \lesssim 3~{\rm GeV}$, where the perturbative QCD is not applicable, and various hadronic decay channels appear (cf., Refs.~\cite{Bauer:2017ris, Aloni:2018vki}).

\begin{figure}[t]
    \centering
    \includegraphics[width=\linewidth]{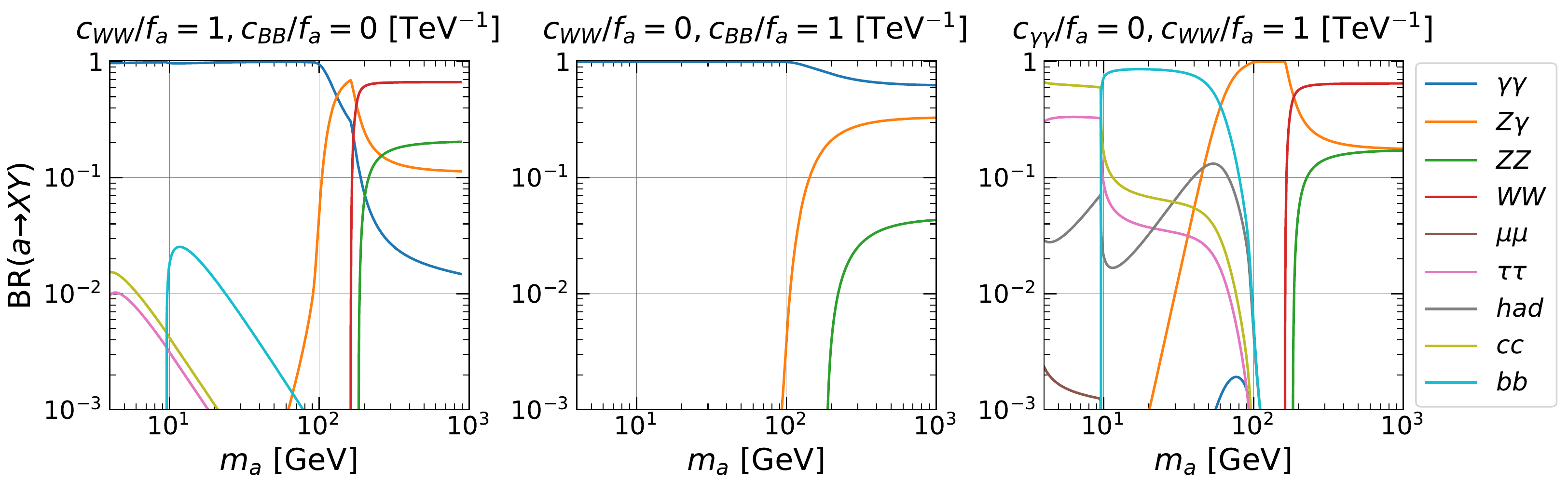}
    \caption{Branching ratios of the ALP with $f_{a}=1\TeV$ and $\Lambda = 4\pi f_{a}$. Here, $c_{BB}=0$, (left) $c_{WW}=0$ (middle), and $c_{\gamma\gamma} \propto g_{a\gamma\gamma} =0$ (right) are set. The decay into light hadrons is denoted as ``had'' in the legend.}
    \label{fig: BRa}
\end{figure}

In Fig.~\ref{fig: BRa}, we show the branching ratios of the ALP with $\Lambda = 4\pi f_{a}$.
Among the ALP couplings, $c_{BB}$ and $c_{WW}$, we set $c_{BB}=0$ on the left panel, $c_{WW}=0$ on the middle, and $c_{\gamma\gamma} \propto g_{a\gamma\gamma} = 0$ on the right.
Here, the decay rate of $a \to Z\gamma$ includes those from the three-body decays, $a \to Z^{*}\gamma\to f\bar f \gamma$. 
In the case with $c_{\gamma\gamma}\neq0$, the decay of $a\to \gamma\gamma$ is dominant for $m_{a}\leq m_{Z}$.
When $c_{BB}=0$, the decay of $a\to WW$ is dominant for $m_{a} > 2m_{W}$, and the decays of $a\to Z\gamma$ and $a\to ZZ$ become relevant above $m_{Z}$ and $2m_{Z}$, respectively.
When $c_{WW}=0$, the decay of $a\to \gamma\gamma$ is still dominant for $m_{a}> m_{Z}$, while the decays of $a\to Z\gamma$ and $a\to ZZ$ become sub-dominant above the thresholds.
In the case with $c_{\gamma\gamma}=0$, the fermionic decay modes are relevant for $m_{a} < m_{Z}$, while the bosonic channels become relevant for $m_{a} \gtrsim m_{Z}$.

Before closing this section, let us comment on the ALP couplings with the $Z$ and Higgs bosons, $g_{aZh}^{\rm eff}$.
It is known that gauge-boson loop contributions to $g_{aZh}^{\rm eff}$ vanish at the one-loop level~\cite{Bauer:2017ris}.
Thus, the ALP decay, $a \to Zh$, is irrelevant for Fig.~\ref{fig: BRa} and the following analysis.

\section{Electroweak precision test (EWPT)} \label{sec: ewpt}

New physics models as well as the SM have been tested by using the EWPOs.
The theoretical predictions are compared with the experimental data by performing global fit analyses.
In particular, the former values are determined within the SM by a set of input observables: the fine-structure constant $\alpha$, the Fermi coupling constant $G_F$, and the $Z$ mass $m_{Z}$. 

The ALP contributions arise in vacuum polarizations due to the ALP interactions with the EW gauge bosons. 
Such contributions have been studied in terms of the oblique parameters, $S$, $T$, and $U$~\cite{Bauer:2017ris}.
It has been known that the ALP generates $S$ and $U$, while $T$ is absent at least at the one-loop level.
Besides, $U$ becomes as large as $S$, in contrast to the case in a wide class of new physics models. 
However, $S$ and $U$ are not enough to represent the ALP contributions. 
We introduce {\it three} additional variables, $\Delta_Z$, $\Delta_W$ and $\Delta\alpha$.\footnote{
No specific parameters are assigned for these variables in Ref.~\cite{Hagiwara:1994pw}.
}
The parameters $\Delta_{Z}$ and $\Delta_{W}$ represent radiative corrections to the gauge coupling constants of the $Z$ and $W$ bosons, respectively.
Also, $\Delta\alpha$ is those to the QED coupling.
Although they are negligible compared to $S$ and $T$ in a wide class of new physics models, this is not the case for the ALP models. 
Moreover, the decay of $Z \to a\gamma$ alters the $Z$-pole observables, such as the hadronic and total decay widths of the $Z$ boson (see Eq.~\eqref{eq: decay_a_VV}).
Since the decay proceeds at the tree level, its impact on EWPOs is likely stronger than those via vacuum polarizations.
We provide their formulae as well as those for $S$, $T$, and $U$ in this section. 

Since the EW data, except for the recent CDF result of the $W$ mass, are consistent with the SM predictions, the ALP contributions should be smaller than the SM ones.
Hence, we retain only the leading contributions from the ALP, which stem from interference terms between the ALP and SM transition amplitudes. 
The ALP contributions are estimated up to the one-loop level.
Besides, since we are interested in the observables at the EW scale, we focus on the leading terms in the $m_f/m_{W}$ expansions, where $m_f$ and $m_{W}$ are the SM fermion and $W$ masses, respectively.

\subsection{Formulation} \label{subsec: Formulation}

In this section, we basically follow Ref.~\cite{Hagiwara:1994pw} for the formulation of the EWPOs.\footnote{
It is also checked that the following results are consistent with the formulae in the literature, \eg, Refs.~\cite{Hollik:1993cg, Hollik:1995dv}. See also, Ref.~\cite{Cirigliano:2013lpa}.
}
We consider the ALP contributions to vacuum polarizations.
They generally appear in the gauge-boson propagators as
\begin{align}
 \Gamma^{ab}_{\mu\nu}(k) = 
 - i g_{\mu\nu} (k^2 - m_{a,0}^2) \delta^{ab} 
 - i \left(g_{\mu\nu} - \frac{k_\mu k_\nu}{k^2} \right) \Pi_T^{ab}(k^2)
 - i \frac{k_\mu k_\nu}{k^2} \Pi_L^{ab}(k^2),
\end{align}
with $a,b=\gamma,W,Z$. 
The mass parameter $m_{a,0}$ is related to the pole mass $m_{a}$ as $m_{a,0}^2 = m_{a}^2 + {\rm Re}\,\Pi_T^{aa}(m_a^2)$.
Here, $\Pi_T^{ab}(k^2)$ and $\Pi_L^{ab}(k^2)$ are the unrenormalized transverse and longitudinal self-energy corrections, respectively.\footnote{
Although the radiative corrections are assumed to include the pinch terms~\cite{Cornwall:1981zr, Cornwall:1989gv, Degrassi:1992ue, Degrassi:1992ff, Binosi:2009qm} in Ref.~\cite{Hagiwara:1994pw}, they are irrelevant in the following ALP analysis because gauge-dependent terms do not arise. 
}
We focus on $\Pi_T^{ab}(k^2)$ because the terms proportional to $k_\mu k_\nu$ are subdominant with respect to the $m_f/m_{W}$ expansion and hereafter neglected. 
For convenience, we define
\begin{align}
 \Pi_{T,V}^{ab}(k^2) = \frac{\Pi_T^{ab}(k^2)-\Pi_T^{ab}(m_{V}^2)}{k^2-m_{V}^2}.
\end{align}
It is also useful to parameterize the vacuum polarizations as
\begin{align}
 \Pi_{T}^{\gamma\gamma}(k^2) &= e^2 \, \Pi_{T}^{QQ}(k^2), \\
 \Pi_{T}^{Z\gamma}(k^2) &= e g_Z \left[ \Pi_{T}^{3Q}(k^2) - s_W^2 \Pi_{T}^{QQ}(k^2) \right], \\
 \Pi_{T}^{ZZ}(k^2) &= g_Z^2 \left[ \Pi_{T}^{33}(k^2) - 2s_W^2 \Pi_{T}^{3Q}(k^2) + s_W^4 \Pi_{T}^{QQ}(k^2) \right], \\
 \Pi_{T}^{WW}(k^2) &= g^2 \, \Pi_{T}^{11}(k^2).
\end{align}

Transition amplitudes for charged currents mediated by the $W$ boson, \eg, $\mu \to e\bar\nu_e\nu_\mu$, are shifted from the tree-level ones as
\begin{align}
   \mathcal{M}^{\rm CC} \to \mathcal{M}^{\rm CC} \left[ 1 - {\rm Re}\,\Pi_{T,W}^{WW}(k^2) \right],
   \label{eq:CCamp}
\end{align}
where $k$ is the momentum carried by the $W$ boson.
Here and hereafter, we ignore imaginary parts of the vacuum polarizations because they correspond to higher-order corrections.
Hence, the extra vacuum-polarization contributions can be implemented by replacing the ${\rm SU(2)}_L$ coupling as
\begin{align}
 g^2 \to \bar g^2(k^2) = g^2 \left[ 1 - {\rm Re}\,\Pi_{T,W}^{WW}(k^2) \right].
 \label{eq: running_g}
\end{align}

Let us next consider neutral currents, $f \bar f \to f'\bar f'$.
Their transition amplitudes are expressed as
\begin{align}
   \mathcal{M}^{\rm NC} = \mathcal{M}^{\rm NC}_{ij} [\bar\psi_f \gamma_\mu P_a \psi_f] [\bar\psi_{f'} \gamma^\mu P_{a'} \psi_{f'}],
\end{align}
where $i = f_a$ and $j = f'_{a'}$ with the chiral projector $P_{a,a'}$.
Similar to the charged currents, the leading contributions from new physics via the vacuum polarizations can be implemented via the effective coupling and weak mixing angles as
\begin{align}
 \mathcal{M}^{\rm NC}_{ij}
 &= 
 \frac{\bar e^2(s)}{s} Q_iQ_j +
 \frac{\bar g_Z^2(s)}{s-m_{Z}^2+is\Gamma_Z/m_{Z}} [I_3^i - Q_i \bar s^2(s)][I_3^j - Q_j \bar s^2(s)],
 \label{eq:NCamp}
\end{align}
where the first term on the right-hand side has a pole at $s=0$, identified as the photon-exchange contribution, and the second one peaks at $s=m_{Z}^2$, corresponding to the $Z$ amplitude.
Here, $s=k^2$ is the squared momentum carried by the photon or $Z$ boson.
The effective parameters are defined as
\begin{align}
 \bar e^2(k^2) &= e^2 \left[ 1 - {\rm Re}\,\Pi_{T,\gamma}^{\gamma\gamma}(k^2) \right], \label{eq: running_e} \\
 \bar g_Z^2(k^2) &= g_Z^2 \left[ 1 - {\rm Re}\,\Pi_{T,Z}^{ZZ}(k^2) \right], \label{eq: running_gZ} \\
 \bar s^2(k^2) &= s_W^2 \left[ 1 + \frac{c_W}{s_W} {\rm Re}\,\Pi_{T,\gamma}^{Z\gamma}(k^2) \right]. 
\end{align}
Since the ALP couples only with the gauge bosons, there are no contributions from vertex and box corrections.

The fine-structure constant, $\alpha$, is associated with the photon-exchange amplitude in the Thomson limit, $k^2\to0$.
The effective coupling at a scale $k^2$ satisfies
\begin{align}
 \frac{1}{\bar \alpha(k^2)} - \frac{1}{\alpha} &= 
 4\pi \, {\rm Re}\! \left[ \Pi_{T,\gamma}^{QQ}(k^2) - \Pi_{T,\gamma}^{QQ}(0) \right],
 \label{eq:alpha}
\end{align}
where $\bar\alpha(k^2) = \bar e^2(k^2)/4\pi$.
Hence, the leading ALP contribution at $k^2 = m_{Z}^2$ gives
\begin{align}
 \bar \alpha(m_{Z}^2) &= 
 \alpha \left\{ 1 - {\rm Re}\left[ \Pi_{T,\gamma}^{\gamma\gamma}(m_{Z}^2) - \Pi_{T,\gamma}^{\gamma\gamma}(0) \right] \right\}
 \equiv 
 \alpha \left( 1 + \Delta\alpha \right).
\end{align}

On the other hand, the measured value of the Fermi coupling constant, $G_F$, is related to the effective coupling of the charged current amplitude at $k^2 \simeq 0$ as
\begin{align}
 G_F = \frac{\bar g^2(0)}{4\sqrt{2}\,m_{W}^2}.
 \label{eq:defGF}
\end{align}
Here, since we focus on radiative corrections from the ALP, vertex and box corrections are ignored. 
It is noted that the $W$ mass is not the input observable, but its theoretical value is determined from Eq.~\eqref{eq:defGF}. 

The effective couplings and weak mixing angle are expressed in terms of $\alpha$, $G_F$, and $m_{Z}$ with the oblique parameters, $S$, $T$, and $U$, as
\begin{align}
 & \frac{1}{\bar g_Z^2(0)} = \frac{1 - \alpha T}{4\sqrt{2}\,G_F m_{Z}^2}, \\
 & \bar s^2(m_{Z}^2) = \frac{1}{2} - \sqrt{\frac{1}{4} - \bar \alpha(m_{Z}^2) \left[ \frac{4\pi}{\bar g_Z^2(0)} + \frac{S}{4} \right]}, \\
 & \frac{4\pi}{\bar g_W^2(0)} = \frac{\bar s^2(m_{Z}^2)}{\bar \alpha(m_{Z}^2)} - \frac{S+U}{4}.
\end{align}
In a given model, the oblique parameters are evaluated as
\begin{align}
 &S = 16\pi \, {\rm Re}\! 
 \left[ \Pi_{T,\gamma}^{3Q}(m_{Z}^2) - \Pi_{T,Z}^{33}(0) \right], \label{eq:S} \\
 &T = \frac{4\sqrt{2}\,G_F}{\alpha} \, {\rm Re}\! \left[ \Pi_{T}^{33}(0) - \Pi_{T}^{11}(0) \right], \label{eq:T} \\
 &U = 16\pi \, {\rm Re}\! \left[ \Pi_{T,Z}^{33}(0) - \Pi_{T,W}^{11}(0) \right].  \label{eq:U}
\end{align}
Thus, the leading contributions from new physics are shown as
\begin{align}
    \bar g_Z^2(0) &= g_Z^2 \left( 1 + \alpha T \right),
    \label{eq: gZsqbar} \\
    \bar s^2(m_{Z}^2) &= 
    s_W^2 \left[
    1 + \frac{c_W^2}{c_W^2-s_W^2} \left( \Delta\alpha - \alpha T \right)
    + \frac{\alpha S}{4 s_W^2 (c_W^2-s_W^2)} \right],
    \label{eq: sWsqbar} \\
    \bar g^2(0) &= 
    g^2 \left[
    1 - \frac{s_W^2 \Delta\alpha}{c_W^2-s_W^2}
    - \frac{\alpha S}{2(c_W^2-s_W^2)} 
    + \frac{c_W^2 \alpha T}{c_W^2-s_W^2}
    + \frac{\alpha U}{4 s_W^2} \right].
    \label{eq: gWsqbar}
\end{align}
It is noticed that the $Z$- and $W$-pole observables, where the gauge bosons carry $k^2 = m_{Z}^2$ and $m_{W}^2$ respectively, require the effective couplings, $\bar g_Z^2(m_{Z}^2)$ and $\bar g^2(m_{W}^2)$.
From Eqs.~\eqref{eq: running_g} and \eqref{eq: running_gZ}, they are related to $\bar g_Z^2(0)$ and $\bar g^2(0)$ as
\begin{align}
 \bar g_Z^2(m_{Z}^2) &= \bar g_Z^2(0) \left\{ 1 - {\rm Re}\left[ \Pi_{T,Z}^{ZZ}(m_{Z}^2) - \Pi_{T,Z}^{ZZ}(0) \right] \right\}
 \equiv 
 g_Z^2(0) \left( 1 + \Delta_Z \right), \\
 \bar g^2(m_{W}^2) &= \bar g^2(0) \left\{ 1 - {\rm Re}\left[ \Pi_{T,W}^{WW}(m_{W}^2) - \Pi_{T,W}^{WW}(0) \right] \right\}
 \equiv 
 g^2(0) \left( 1 + \Delta_W \right).
\end{align}
Consequently, the EWPOs are evaluated theoretically in terms of $\Delta_Z$, $\Delta_W$, and $\Delta\alpha$ as well as $S$, $T$, and $U$ (see also Sec.~\ref{sec: observables}).

\subsection{ALP contributions} \label{sec: ALP_contribution}

In Eq.~\eqref{eq: effective_coupling}, the ALP couples with $\gamma$, $Z$, and $W$. 
In the $R_{\xi}$ gauge, its contributions to the vacuum polarizations at the one-loop level are derived as
\begin{align}
    \Pi_T^{WW}(k^2) &= 
    \frac{1}{288\pi^2}g_{aWW}^{2}F(k^2;a,W), 
    \\
    \Pi_T^{\gamma\gamma}(k^2) &= 
    \frac{1}{288\pi^2}
    \left[g_{a\gamma\gamma}^2 F(k^2;a,\gamma) + g_{aZ\gamma}^2 F(k^2;a,Z) \right],
    \\
    \Pi_T^{Z\gamma}(k^2) &= 
    \frac{1}{288\pi^2}
    \left[g_{a\gamma\gamma}g_{aZ\gamma} F(k^2;a,\gamma) + g_{aZ\gamma}g_{aZZ} F(k^2;a,Z) \right],
    \\
    \Pi_T^{ZZ}(k^2) &= 
    \frac{1}{288\pi^2}
    \left[g_{aZ\gamma}^2 F(k^2;a,\gamma) + g_{aZZ}^2 F(k^2;a,Z) \right].
\end{align}
These results are valid for any ALP mass.
The loop function is defined as
\begin{align}
 F(k^2;a, V) &= 
 3 \left[k^2 - (m_a + m_{V})^2\right] \left[k^2 - (m_a - m_{V})^2\right] 
 \notag \\ &\quad\quad 
 \times \left[ B_0(k^2;m_a,m_{V}) - B_0(0;m_a,m_{V}) \right]
 \notag \\ &\quad 
 - 3 k^2 \left[ A_0(m_a) + A_0(m_{V}) + (2m_a^2 + 2m_{V}^2 - k^2) B_0(0;m_a,m_{V}) \right]
 \notag \\ &\quad
 + 7 k^2 (3m_a^2 + 3m_{V}^2 - k^2), \label{eq: F_function}
\end{align}
where $A_0(x)$ and $B_0(k^{2}; x, y)$ are the Passarino-Veltman functions~\cite{Passarino:1978jh} and shown explicitly in Appendix~\ref{app: PV_funcs}.
It is noticed that the results are independent of the gauge parameters. 
Then, the ALP contributions to $S$, $T$, and $U$ are given by
\begin{align}
    \alpha S &=
        \frac{2c_{W}^{2}s_{W}^{2}}{9\pi^{2}m_{Z}^{2}}
        \frac{c_{WW}c_{BB}}{f_{a}^{2}}
        \qty[F(m_{Z}^{2}; a, \gamma)-F(m_{Z}^{2}; a, Z)], \\
    \alpha T &= 0, \\
    \alpha U &=
        \frac{2s_{W}^{4}}{9\pi^{2}m_{Z}^{2}}
        \frac{c_{WW}^{2}}{f_{a}^{2}}
        \qty[F(m_{Z}^{2}; a, \gamma)+\frac{c_{W}^{2}}{s_{W}^{2}}F(m_{Z}^{2}; a, Z)-\frac{1}{s_{W}^{2}c_{W}^{2}}F(m_{W}^{2}; a, W)].
\end{align}
It is noticed that the ALP does not contribute to $T$.
On the other hand, the contributions to $\Delta\alpha$, $\Delta_{Z}$, and $\Delta_{W}$ are given by
\begin{align}
    \Delta\alpha &=
        -\frac{g_{a\gamma\gamma}^{2}}{288\pi^{2}}\qty[\frac{F(m_{Z}^{2};a,\gamma)}{m_{Z}^{2}}-F^{\prime}(0; a,\gamma)]
        -\frac{g_{aZ\gamma}^{2}}{288\pi^{2}}\qty[\frac{F(m_{Z}^{2};a,Z)}{m_{Z}^{2}}-F^{\prime}(0; a,Z)], 
        \label{eq: delta_alpha_limit} \\
    \Delta_{Z} &= 
        \frac{g_{aZ\gamma}^{2}}{288\pi^{2}}\qty[\frac{F(m_{Z}^{2};a,\gamma)}{m_{Z}^{2}}-F^{\prime}(m_{Z}^{2}; a,\gamma)]
        +\frac{g_{aZZ}^{2}}{288\pi^{2}}\qty[\frac{F(m_{Z}^{2};a,Z)}{m_{Z}^{2}}-F^{\prime}(m_{Z}^{2}; a,Z)], \\
    \Delta_{W} &= \frac{1}{288\pi^{2}}g_{aWW}^{2}\qty[\frac{F(m_{W}^{2};a,W)}{m_{W}^{2}}-F^{\prime}(m_{W}^{2};a,W)],
    \label{eq: delta_W_limit}
\end{align}
where $F^{\prime}(k^{2}; a, V)$ is a derivative of $F(k^{2}; a, V)$ and expressed explicitly as
\begin{align}
F^{\prime}(k^2;a, V) &= 
6 \left(k^2 - m_a^2 - m_{V}^2\right)
 \left[ B_0(k^2;m_a,m_{V}) - B_0(0;m_a,m_{V}) \right]
 \notag \\ &\quad
 +3 \left[k^2 - (m_a + m_{V})^2\right] \left[k^2 - (m_a - m_{V})^2\right]B_0^{\prime}(k^2;m_a,m_{V})
  \notag \\ &\quad
 - 3\left[ A_0(m_a) + A_0(m_{V}) + (2m_a^2 + 2m_{V}^2 - k^2) B_0(0;m_a,m_{V}) \right]
 \notag \\ &\quad
 + 7(3m_a^2 + 3m_{V}^2 - 2k^2).
\end{align}
The relation between $(c_{BB}, c_{WW})$ and $(g_{a\gamma\gamma}, g_{aZ\gamma}, g_{aZZ})$ is found in Eqs.~\eqref{eq: gaAA}--\eqref{eq: gaZZ}.
It is noticed that $\Delta\alpha$, $\Delta_{Z}$, and $\Delta_{W}$ can be as large as $\alpha S$ and $\alpha U$, and thus, must not be neglected.

Let us show the above results in the light ALP-mass limit, $m_{a}\ll m_{V}\, (V=Z,W)$.
They are obtained as
\begin{align}
    \alpha S &=
        -\frac{2c_{W}^{2}s_{W}^{2}m_{Z}^{2}}{\pi^{2}}
        \frac{c_{WW}c_{BB}}{f_{a}^{2}}
        \qty(\ln{\frac{m_{Z}^{2}}{\Lambda^{2}}}+1),
        \\
    \alpha U &=
        -\frac{2s_{W}^{4}m_{Z}^{2}}{3\pi^{2}}
        \frac{c_{WW}^{2}}{f_{a}^{2}}
        \qty(\ln{\frac{m_{Z}^{2}}{\Lambda^{2}}}+\frac{1}{3}+\frac{2c_{W}^{2}}{s_{W}^{2}}\ln{\frac{m_{W}^{2}}{m_{Z}^{2}}}), \\
    \Delta \alpha &=
        \frac{m_{Z}^{2}}{96\pi^{2}}\qty[
            g_{a\gamma\gamma}^{2}\qty(\ln{\frac{m_{Z}^{2}}{\Lambda^{2}}}+\frac{11}{3})
            +g_{aZ\gamma}^{2}\qty(\ln{\frac{m_{Z}^{2}}{\Lambda^{2}}}+\frac{11}{6})], 
            \\
    \Delta_{Z} &= \frac{m_{Z}^{2}}{96\pi^{2}}\qty(g_{aZ\gamma}^{2}+g_{aZZ}^{2})\qty(\ln{\frac{m_{Z}^{2}}{\Lambda^{2}}}+\frac{4}{3}), 
    \\
    \Delta_{W} &= \frac{m_{W}^{2}}{96\pi^{2}}g_{aWW}^{2}\qty(\ln{\frac{m_{W}^{2}}{\Lambda^{2}}}+\frac{4}{3}). 
\end{align}
The results for $S$, $T$, $U$, and $\Delta \alpha$ are consistent with those in Ref.~\cite{Bauer:2017ris}.

Let us comment on the ALP contributions, $U$, $\Delta\alpha$, $\Delta_{Z}$, and $\Delta_{W}$.
In a wide class of new physics models in a high-energy scale, they are suppressed compared to $S$ and $T$. 
However, this is not the case for the ALP models. 
This is because the ALP contributions to the vacuum polarizations vanish at $k^{2}=0$ (see Eq.~\eqref{eq: F_function}).
Then, the leading contribution to $S$ arises from the first-order derivative of the vacuum polarization, which is comparable to $U$, $\Delta\alpha$, $\Delta_{Z}$, and $\Delta_{W}$.

\subsection{Observables} \label{sec: observables}

The ALP contributions to the $Z$-pole observables are evaluated by the last term of the transition amplitude in Eq.~\eqref{eq:NCamp}.
It is noticed that this term has the same form as the tree-level one except that the coupling and weak mixing are replaced by the effective ones. 
Hence, the ALP contributions are represented in terms of the following effective charges of the $Z$ coupling,
\begin{align}
 g_{V,f} &= \sqrt{\rho_Z} \left[ I_{3}^f - 2\, Q_f \bar s^2(m_{Z}^2) \right]
 \equiv \hat g_{V,f} + \Delta g_{V, f}, \\
 g_{A,f} &= \sqrt{\rho_Z} \, I_{3}^f
 \equiv \hat g_{A,f} + \Delta g_{A, f},
\end{align}
with $\hat g_{V,f} = I_{3}^f - 2\, Q_f s_W^2$ and $\hat g_{A,f} = I_{3}^f$.
Note that the effective parameters should be estimated at $k^2=m_{Z}^2$.
Here, $\rho_Z$ is defined as
\begin{align}
 \rho_Z = \bar g_Z^2(m_{Z}^2)/g_Z^2 = 1 + \alpha T + \Delta_Z. 
\end{align}
Here and hereafter, variables with a hat $(\,\hat{}\,)$, \eg, $\hat g_{V,f}$ and $\hat g_{A,f}$, represent the SM contributions at the tree level, and those with $\Delta$ such as $\Delta g_{V, f}$ and $\Delta g_{A, f}$ denote the ALP corrections. 
On the other hand, those with a subscript ``SM'' show the SM predictions including radiative corrections.

When we evaluate the EWPOs, the ALP corrections are dominated by interference terms between the ALP and SM transition amplitudes. 
In particular, it is sufficient for the latter to be evaluated at the tree level when we focus on the leading ALP contributions.
Hence, by expressing the partial decay width for $Z\to f\bar{f}$ as
\begin{align}
 \Gamma_f = (\Gamma_f)_{\rm SM} + \Delta \Gamma_f,
\end{align}
the SM contribution at the tree level, $\hat \Gamma_f$, is given by
\begin{align}
 \hat \Gamma_f = 
 N_C^f \frac{G_F m_{Z}^3}{6\sqrt{2}\pi} \left[ \left(\hat g_{V,f}\right)^2 + \left(\hat g_{A,f}\right)^2 \right],
\end{align}
and $\Delta \Gamma_f$ is obtained as 
\begin{align}
 \Delta \Gamma_f = 
 N_C^f \frac{G_F m_{Z}^3}{3\sqrt{2}\pi} \left[ \left(\hat g_{V,f}\right)^2 + \left(\hat g_{A,f}\right)^2 \right]
 \qty(\frac{\Delta g_{V, f}}{\hat g_{V, f}}+\frac{\Delta g_{A, f}}{\hat g_{A, f}}).
\end{align}
The radiative corrections to the SM value will be mentioned in Sec.~\ref{sec: analysis_strategy}.
 
When $m_{a}$ is smaller than $m_{Z}$, the decay channel $Z \to a \gamma$ is open.
Its partial decay width is given by
\begin{align}
  \Gamma_{a\gamma} \equiv \Gamma(Z \to a \gamma) = \frac{m_{Z}^{3}}{96\pi}g_{aZ\gamma}^{2}\qty(1-\frac{m_{a}^{2}}{m_{Z}^{2}})^{3}.
\end{align}
Then, the hadronic and total decay widths of the $Z$ boson are modified as
\begin{align}
 \Gamma_{\rm had} &= \Gamma_u + \Gamma_d + \Gamma_c + \Gamma_s + \Gamma_b
 \equiv (\Gamma_{\rm had})_{\rm SM} + \Delta\Gamma_{\rm had}, \\
 \Gamma_Z &= \Gamma_e + \Gamma_\mu + \Gamma_\tau + 3 \Gamma_\nu + \Gamma_{\rm had} + \Gamma_{a\gamma}
 \equiv (\Gamma_Z)_{\rm SM} + \Delta\Gamma_Z,
\end{align}
where $\Delta\Gamma_{{\rm had}, Z}$ are the sums of $\Delta \Gamma_f$.
It is noticed that $\Gamma_{a\gamma}$ appears in $\Gamma_Z$ and is generated at the tree level with $g_{aZ\gamma}$. 
Hence, its impact on the EWPO global fit is likely stronger than those via vacuum polarizations. 

The other $Z$-pole observables relevant for the following analysis are represented explicitly as follows. 
First of all, the hadronic cross-section is shown as
\begin{align}
    \sigma_{\rm had}^0 &= (\sigma_{\rm had}^0)_{\rm SM}
    + \frac{12\pi}{m_{Z}^2} \frac{\hat\Gamma_{e}\hat\Gamma_{\rm had}}{\hat\Gamma_{Z}^2}
    \qty(
        \frac{\Delta \Gamma_{e}}{\hat\Gamma_{e}}
        +\frac{\Delta \Gamma_{\rm had}}{\hat\Gamma_{\rm had}}
        -2\frac{\Delta \Gamma_{Z}}{\hat\Gamma_{Z}}
    ).
\end{align}
Next, the ratios of the partial decay widths become
\begin{align}
    R^0_\ell &= (R^0_\ell)_{\rm SM}
    + \frac{\hat\Gamma_{\rm had}}{\hat\Gamma_{\ell}}
    \qty(
        \frac{\Delta \Gamma_{\rm had}}{\hat\Gamma_{\rm had}}
        -\frac{\Delta \Gamma_{\ell}}{\hat\Gamma_{\ell}}
    ), \\
    R^0_q &= (R^0_q)_{\rm SM}
    + \frac{\hat\Gamma_{q}}{\hat\Gamma_{\rm had}} 
    \qty(
        \frac{\Delta \Gamma_{q}}{\hat\Gamma_{q}}
        -\frac{\Delta \Gamma_{\rm had}}{\hat\Gamma_{\rm had}}
    ),
\end{align}
where $\ell$ denotes the SM leptons. 
Also, $\sin^2\theta^f_{\rm eff}$ is given by
\begin{align}
 \sin^2\theta^f_{\rm eff} &= (\sin^2\theta^f_{\rm eff})_{\rm SM}
 -\frac{1}{4|Q_f|} \frac{\hat g_{V,f}}{\hat g_{A,f}}
 \left( \frac{\Delta g_{V,f}}{\hat g_{V,f}} -\frac{\Delta g_{A,f}}{\hat g_{A,f}}\right).
\end{align}
The left-right asymmetry is shown as
\begin{align}
    \mathcal{A}_f &= (\mathcal{A}_f)_{\rm SM}
    +\mathcal{\hat A}_f
    \qty(1-\frac{\hat g_{V,f}}{\hat g_{A,f}}\mathcal{\hat A}_f)
    \qty( \frac{\Delta g_{V,f}}{\hat g_{V,f}} -\frac{\Delta g_{A,f}}{\hat g_{A,f}}).
\end{align}
Here, the SM contribution at the tree level is
\begin{align}
    \mathcal{\hat A}_f = \frac{2 \hat g_{V,f}/\hat g_{A,f}}{1 + (\hat g_{V,f}/\hat g_{A,f})^2}.
\end{align}
Finally, the forward-backward asymmetry is given by
\begin{align}
    A^0_{\rm FB} = 
    (A^0_{\rm FB})_{\rm SM}
    + \frac{3}{4}\mathcal{\hat A}_e\mathcal{\hat  A}_f
    \qty(
        \frac{\Delta \mathcal{A}_e}{\mathcal{\hat A}_e}
        +\frac{\Delta \mathcal{A}_f}{\mathcal{\hat A}_f}
    ).
\end{align}

On the other hand, the mass and decay widths of the $W$ boson are also evaluated theoretically. 
From Eqs.~\eqref{eq:defGF} and \eqref{eq: gWsqbar}, the mass is obtained as
\begin{align}
    m_{W}^2
    &=
    (m_{W}^2)_{\rm SM} 
    + \frac{\alpha c_W^2 m_{Z}^2}{c_W^2-s_W^2}
    \left[ 
    - s_W^2 \frac{\Delta\alpha}{\alpha}
    - \frac{S}{2} 
    + c_W^2 T
    + \frac{c_W^2-s_W^2}{4 s_W^2} U
    \right]
    \notag \\
    &\equiv 
    (m_{W}^2)_{\rm SM} + \Delta m_{W}^2.
\end{align}
Then, the $W$ partial decay width becomes
\begin{align}
    \Gamma(W \to f_i f_j)
    &=
    [(\Gamma_W)_{ij}]_{\rm SM}
    + (\hat\Gamma_W)_{ij}
    \bigg[ 
    \frac{3\Delta m_{W}^2}{2c_W^2 m_{Z}^2} + \Delta_W
    \bigg].
\end{align}
Here, $\Delta_W$ arises because the $W$ coupling is estimated at $k^2 = m_{W}^2$.
The SM prediction at the tree level is expressed as
\begin{align}
 (\hat\Gamma_W)_{ij} &=
 N_c^f |V_{ij}|^2 \frac{g^2}{48\pi} m_{W}.
\end{align}
where $V_{ij}$ is the CKM matrix when $i,j$ are quarks, while it is the identity matrix for leptons.

\subsection{Analysis strategy} \label{sec: analysis_strategy}

\begin{table}[t]
\centering
\begin{tabular}{ccc|ccc}
\hline
& Measurement & Ref. &
& Measurement & Ref.
\\
\hline
$\alpha_s(m_{Z}^2)$ & 
$0.1177 \pm 0.0010$ &
\cite{deBlas:2022hdk} &
$m_{Z}$ [GeV] &
$91.1875 \pm 0.0021$ &
\cite{Janot:2019oyi} 
\\
\cline{1-3}
$\Delta\alpha_{\mathrm{had}}^{(5)} (m_{Z}^2)$ &
$0.02766 \pm 0.00010$ &
\cite{deBlas:2022hdk} &
$\Gamma_Z$ [GeV] &
$2.4955 \pm 0.0023$ &
\\
\cline{1-3}
$m_t$ [GeV] &
$172.69 \pm 0.30$ &
\cite{ParticleDataGroup:2022pth} &
$\sigma_{h}^{0}$ [nb] &
$41.4802 \pm 0.0325$ &
\\
\cline{1-3}
$m_h$ [GeV] &
$125.21 \pm 0.17$ &
\cite{ParticleDataGroup:2022pth} &
$R^{0}_{\ell}$ &
$20.7666 \pm 0.0247$ &
\\
\cline{1-3}
$m_{W}$ [GeV] &
$80.377 \pm 0.012$ &
\cite{ParticleDataGroup:2022pth} &
$A_{\mathrm{FB}}^{0, \ell}$ &
$0.0171 \pm 0.0010$ &
\\
\cline{4-6}
&
$80.4133 \pm 0.0080$ &
\cite{deBlas:2022hdk} &
$R^{0}_{b}$ &  
$0.21629 \pm 0.00066$ & 
\cite{ALEPH:2005ab,Bernreuther:2016ccf}
\\
\cline{1-3}
$\Gamma_{W}$ [GeV] & 
$2.085 \pm 0.042$ &
\cite{ParticleDataGroup:2022pth} &
$R^{0}_{c}$ & 
$0.1721 \pm 0.0030$ & 
\\
\cline{1-3}
$\mathcal{B}(W\to \ell\nu)$ &
$0.10860 \pm 0.00090$ &
\cite{Schael:2013ita} &
$A_{\mathrm{FB}}^{0, b}$ & 
$0.0996 \pm 0.0016$ &
\\
\cline{1-3}
$\mathcal{A}_{\ell}$ (LEP) & 
$ 0.1465 \pm 0.0033 $ &
\cite{ALEPH:2005ab} &
$A_{\mathrm{FB}}^{0, c}$ & 
$0.0707 \pm 0.0035$ &
\\
$\mathcal{A}_{\ell}$ (SLD) & 
$ 0.1513 \pm 0.0021 $ &
\cite{ALEPH:2005ab} &
$\mathcal{A}_b$ & 
$0.923 \pm 0.020$ &
\\
& 
&
&
$\mathcal{A}_c$ & 
$0.670 \pm 0.027$ &
\\
\hline
\end{tabular}
\caption{Experimental data of the SM input parameters and EWPOs.}
\label{tab:EWPO}
\end{table}

As a test of the ALP models, we perform a global fit analysis on the EWPOs. 
A likelihood function is defined by a multivariate Gaussian, $-2 \ln L = (\vb{y}-\vb*{\mu})^{T}V^{-1}(\vb{y}-\vb*{\mu})$, where $\vb{y}$ is a vector of the measured quantities, $\vb*{\mu}$ is the corresponding theoretical predictions, and $V$ is the covariance matrix.

In Table~\ref{tab:EWPO}, we summarize the measured values of the SM input and the EWPOs.
The covariance matrices are provided in the references for those listed in the right column.
The other SM input parameters, such as $G_{F}$, $\alpha$, and the masses of the light SM fermions, are fixed to be the observed central values~\cite{ParticleDataGroup:2022pth}.

In the table, we show two results of the $W$ mass $m_{W}$; the one provided by Particle Data Group (PDG)~\cite{ParticleDataGroup:2022pth} and a value combined with the recent CDF result, for which we adopt the averaged value in Ref.~\cite{deBlas:2022hdk}. 
They will be denoted by $m_{W}^{\rm PDG}$ and $m_{W}^{\rm CDF}$, respectively.
In the following analysis, we study both cases in the ALP models.
On the other hand, the SM prediction of $m_{W}$ has been evaluated at the two-loop level~\cite{Awramik:2003rn}.
We also include theoretical uncertainties from unknown higher-order corrections, $\delta m_{W}=0\pm4\MeV$ (cf., Ref.\cite{deBlas:2022hdk}).
As a result, based on the SM input in the table, the SM prediction is derived as
\begin{align}
  (m_{W})_{\rm SM} = 80.3552 \pm 0.0055\GeV.
\end{align}
The result is consistent with the PDG value, $m_{W}^{\rm PDG}$, but deviates from the result including the CDF result, $m_{W}^{\rm CDF}$, at the $6\sigma$ level.\footnote{
The largest uncertainty of the SM prediction originates in $m_{Z}$. 
Although the uncertainty from the top quark mass seems to be smaller, it may involve a potentially larger (theoretical) uncertainty, which reduces the discrepancy (cf., Ref.~\cite{deBlas:2022hdk}). 
}

The SM predictions for the $Z$-pole observables as well as the $W$ mass have been evaluated up to the two-loop level including the electroweak corrections~\cite{Awramik:2006uz, Dubovyk:2019szj}.
On the other hand, for the $W$ decay widths, we use the SM predictions provided in Ref.~\cite{dEnterria:2020cpv}. In particular, those assuming the CKM unitarity are adopted.
Theoretical uncertainties from unknown higher-order corrections are irrelevant for those observables and neglected in the following analysis unlikely to $m_{W}$.

We implement the ALP contributions to the EWPOs, as explained in Sec.~\ref{sec: ALP_contribution}.
Obviously, $\Delta\alpha$, $\Delta_{Z}$, $\Delta_{W}$ and $\Gamma_{a\gamma}$ must be taken into account at the same as $S$, $T$, and $U$ when we perform the EWPT.
This feature is contrary to the analyses explored in the previous works~\cite{Bauer:2017ris, Bauer:2018uxu, Yuan:2022cpw}, where $\Delta\alpha$, $\Delta_{Z}$, $\Delta_{W}$ and $\Gamma_{a\gamma}$ were ignored in the analyses of $S$ and $U$.
In particular, they focused on the ALP much lighter than the $Z$ boson; in such a case $\Gamma_{a\gamma}$ alters $\Gamma_Z$ drastically.
We will show the numerical results in Sec.~\ref{sec: result}.

\section{Experimental constraints} \label{sec: constraint}

In this section, we summarize experimental constraints on the ALP models.
Although there are very severe bounds from cosmological measurements~\cite{Jaeckel:2010ni, Cadamuro:2011fd, Proceedings:2012ulb}, the model can avoid them if the ALP mass is $m_a \gtrsim 1\GeV$.
Then, constraints from flavor and collider experiments become relevant. 

\subsection{Flavor constraints}

Let us first consider the flavor constraints.
If the ALPs have an interaction with the $W$ boson, they generate quark-flavor transitions via $W$ loops (see Eq.~\eqref{eq: flavor_violating _coupling})~\cite{Izaguirre:2016dfi, Alonso-Alvarez:2018irt, Gavela:2019wzg, Guerrera:2021yss, Bauer:2021mvw, Guerrera:2022ykl}.
In particular, for $m_a = \mathcal{O}(1)\GeV$, the $B$-meson decays provide the best sensitivities (see, \eg, Ref.~\cite{Bauer:2021mvw}). 
The decay rate for $B^{+}\to K^{+}a$ is obtained as~\cite{Izaguirre:2016dfi, Gavela:2019wzg, Bauer:2021mvw, Guerrera:2022ykl}
\begin{align}
    \Gamma(B^{+}\to K^{+}a) = \frac{m_{B}^{3}}{64\pi}\abs{g_{abs}^{\rm eff}}^{2}f_{0}(m_{a}^{2})\lambda_{Ka}^{1/2}\qty(1-\frac{m_{K}^{2}}{m_{B}^{2}}),
\end{align}
where $\lambda_{Ka} = \lambda(m_a^2/m_B^2,m_K^2/m_B^2)$ with the $B$ $(K)$ meson mass, $m_B$ $(m_K)$.
Here, $f_{0}(q^{2})$ is the scalar form factor, which is evaluated by following Ref.~\cite{FlavourLatticeAveragingGroupFLAG:2021npn}. 

In the ALP parameter region in interest, the ALP produced from the meson is likely to decay within the detectors.
If its lifetime is short enough, it decays promptly.
As the lifetime increases, the ALP propagates inside the detectors before it decays.
Then, the vertex constructed from the final-state particles of the ALP decay becomes displaced from the primary interaction vertex. 
Among various such channels (see \eg, Ref.~\cite{Bauer:2021mvw}), the following two studies provide the best sensitivities:
\begin{itemize}
\item When $g_{a\gamma\gamma}$ is large enough, the ALP decays predominantly into a pair of photons. 
The severest constraint is obtained from $B^+ \to K^+ a$, $a\to\gamma\gamma$ in the mass range of $0.175 < m_a < 4.78\GeV$ with various lifetimes.
The analysis was performed by the BaBar collabotation~\cite{BaBar:2021ich}. 
\item When $g_{a\gamma\gamma}$ is suppressed, the ALP decays into SM fermions (see Sec.~\ref{sec: decay}).
Among the decay channels, the process $B^+ \to K^+ a$, $a\to\mu\mu$, though ${\rm Br}(a\to\mu\mu)$ is $\sim 10^{-3}$, provides a severe bound for $m_a = \mathcal{O}(1)\GeV$.
The analysis was performed by the LHCb collaboration in the mass range of $0.25<m_a<4.7\GeV$ and the lifetime $0.1<\tau_a<1000\,{\rm ps}$~\cite{LHCb:2016awg}.
\end{itemize}
These constraints are very severe and inevitable as long as the $B$-meson decays into the ALP, \ie, for $m_a < m_B-m_K$.\footnote{
There are several mass regions in which the SM backgrounds are huge and the flavor constraints become absent or relaxed drastically. 
See Refs.~\cite{BaBar:2021ich, LHCb:2016awg} for the details. 
}
Thus, the ALPs are favored to be heavier than $4.8\GeV$ to avoid the constraints and contribute to the EWPOs effectively.

\subsection{Collider constraints}

Even when the ALP is heavier than $4.8\GeV$ and the flavor constraints are avoided, the model is subject to the collider constraints. 
In the mass range of $m_a \gtrsim 1\GeV$, the ALP productions followed by $a\to\gamma\gamma$ have been studied with the experimental data of LEP~\cite{Mimasu:2014nea, Jaeckel:2015jla, Knapen:2016moh}, Belle II~\cite{Belle-II:2020jti}, CDF~\cite{Mimasu:2014nea, CDF:2013lma}, LHC (the proton collisions)~\cite{Jaeckel:2012yz, Bauer:2017ris, Bauer:2018uxu, Florez:2021zoo, dEnterria:2021ljz, Wang:2021uyb}, and LHC (the Pb collisions)~\cite{Knapen:2016moh, CMS:2018erd, ATLAS:2020hii}. 
There are also studies focusing on the ALP interaction with the $W$ bosons~\cite{Craig:2018kne, Bonilla:2022pxu}. 

Let us first consider the case that the produced ALP decays into a pair of photons. 
The collider constraints are summarized, \eg, in Ref.~\cite{dEnterria:2021ljz}.
For $1 \lesssim m_a \lesssim 5\GeV$, the process $e^+e^- \to \gamma a \to 3\gamma$ provides the best sensitivity~\cite{Mimasu:2014nea, Jaeckel:2015jla, Knapen:2016moh, Belle-II:2020jti}. 
The scattering cross section of $e^+e^- \to \gamma a$ is expressed as~\cite{Bauer:2017ris}
\begin{align}
 \frac{d\sigma(e^+e^-\to\gamma a)}{d\Omega} 
 = \frac{\alpha}{128\pi}s^2\qty(1-\frac{m_a^2}{s})^3 (1+\cos^2\theta) \qty(\abs{V}^2 + \abs{A}^2),
 \label{eq: ee_to_agamma}
\end{align}
where $\theta$ is the photon angle relative to the beam direction, and $V,\ A$ are
\begin{align}
 V &= \frac{g_{a\gamma\gamma}}{s}+\frac{1-4s_W^2}{4c_Ws_W}\frac{g_{aZ\gamma}}{s-m_{Z}^2+is\Gamma_Z/m_{Z}}, \\
 A &= \frac{1}{4c_Ws_W}\frac{g_{aZ\gamma}}{s-m_{Z}^2+is\Gamma_Z/m_{Z}}.
\end{align}
The first term on the right-hand side of $V$ corresponds to a photon-exchange contribution, and the second one and $A$ to the $Z$ boson. 
The LEP and LEP II experiments have measured $e^+e^- \to \gamma\gamma (\gamma)$ at the center-or-energy around the $Z$ pole and $200\GeV$, respectively.
The results have been used to constraint the ALP:
\begin{itemize}
\item 
The former case, \ie, for $\sqrt{s} \sim m_{Z}$, was analyzed by Ref.~\cite{Jaeckel:2015jla}.
Since the on-shell $Z$ boson contributions dominate the cross section, the results provided by Ref.~\cite{Jaeckel:2015jla} can be interpreted as a bound on $\Gamma(Z\to \gamma a) \times {\rm Br}(a \to \gamma\gamma)$.\footnote{
Although the ATLAS collaboration reported a stronger bound on ${\rm BR}(Z\to3\gamma)$~\cite{ATLAS:2015rsn}, the result is applied to a higher $m_a$ region because of photon kinematical cuts~\cite{Bauer:2017ris}.
}

On the other hand, when $g_{aZ\gamma}$ is suppressed, the scattering proceeds by exchanging an off-shell photon. 
Since the cross section is small, its constraint is sufficiently weak for the parameter region in interest. 
\item
The LEP II data in the latter case, \ie, for $\sqrt{s} \sim 200\GeV$, were analyzed by Ref.~\cite{Knapen:2016moh}.
In the above mass range, since the photons from the ALP decay are likely to be collimated, the inclusive $e^+e^- \to 2\gamma$ signal regions were studied. 
Although the scattering was supposed to proceed only via an off-shell photon in the literature, the $Z$ boson contributes generally as well. 
It is noticed from Eq.~\eqref{eq: ee_to_agamma} that the latter contribution does not affect the kinematic distributions of the final-state particles. 
Thus, we rescale the constraints provided in Ref.~\cite{Knapen:2016moh} by $\sigma(e^+e^-\to\gamma a) \times {\rm Br}(a \to \gamma\gamma)$. 
\end{itemize}

The LHC results~\cite{Jaeckel:2012yz, Bauer:2017ris, Bauer:2018uxu, Florez:2021zoo, dEnterria:2021ljz, Wang:2021uyb, Knapen:2016moh, CMS:2018erd, ATLAS:2020hii} give the best sensitivity for $m_a \gtrsim 5\GeV$, where the ALPs are assumed to be produced via vector gauge-boson fusions (VBFs):
\begin{itemize}
\item 
Resonant productions of the ALPs from photon fusions, $\gamma\gamma \to a \to \gamma\gamma$, have been studied in the Pb-Pb collision at LHC~\cite{Knapen:2016moh, CMS:2018erd, ATLAS:2020hii}.
The initial photons are emitted from the Pb nuclei. 
Heavy-ion collisions have been considered because the electric charge is larger than the proton. 
This process provides the best sensitivity for $5 \lesssim m_a \lesssim 100\GeV$.
\item 
For $100 \lesssim m_a \lesssim 160\GeV$, resonant productions of the ALPs via VBFs, $VV \to a \to \gamma\gamma$ $(V=\gamma, Z, W)$, have been studied by using the ATLAS analyses on Higgs boson productions via VBFs
in the p-p collisions~\cite{Jaeckel:2012yz}.
Here, the initial gauge bosons are emitted from quarks in the incoming protons.
\item 
For $150 \lesssim m_a \lesssim 400\GeV$, VBF productions of the ALPs in the p-p collisions contribute to the signals~\cite{Jaeckel:2012yz}.
It is found in the literature that a large kinematical cut is imposed on the invariant mass of the final-state photons, and their constraints can be interpreted as those on the non-resonant productions of the ALPs via photon fusions, $\gamma\gamma \to a^* \to \gamma\gamma$, where the initial photons are emitted from the proton.
\item 
More recently, the reference~\cite{Bauer:2018uxu} has derived a bound on $\gamma\gamma \to a \to \gamma\gamma$ by recasting the ATLAS result on the search for spin-0 resonances~\cite{ATLAS:2017ayi}. 
The result shows the best sensitivity for $200 \lesssim m_a < 2700\GeV$.
\end{itemize}
The constraints from the on-shell productions of the ALPs via $\gamma\gamma \to a \to \gamma\gamma$ are interpreted as those on $\abs{g_{a\gamma\gamma}^{\rm eff}}^2{\rm Br}(a\to \gamma\gamma)$.
On the other hand, those from the non-resonant productions, $\gamma\gamma \to a^* \to \gamma\gamma$, are recast by $\abs{g_{a\gamma\gamma}^{\rm eff}}^4$.

Next, let us consider the collider constraints for the ALPs decaying into particles other than a pair of photons. 
They become significant especially when $g_{a\gamma\gamma}$ is absent at the tree level.
The LEP and LHC bounds, in this case, have been studied in Ref.~\cite{Craig:2018kne}, where the ALPs are assumed to be produced at the on-shell. 
The most relevant ones on each mass region are summarized as follows:
\begin{itemize}
\item 
For $1 \lesssim m_a \lesssim 40\GeV$, the leading constraint is from $Z \to \gamma a \to \gamma jj$, where $j$ denotes a jet (gluon and quarks).
The decay was measured on the $Z$ pole at the LEP experiment.
The bounds are understood as those on $\Gamma(Z \to \gamma a) \times \sum_{j}{\rm Br}(a \to jj)$ with $j = {\rm light\ hadrons}, c, b$. 
\item
For $40\GeV \lesssim m_a \lesssim m_{Z}$, the search for $Z \to \gamma a$, $a \to \nu\bar\nu \gamma$ at LEP gives a significant constraint. 
Here, the neutrinos are produced by exchanging an off-shell $Z$ boson, \ie, $a \to Z^* \gamma \to \nu\bar\nu \gamma$.
Such a three-body decay can compete with $a \to ff$ because the former (latter) proceeds at the tree level (via radiative corrections).
The bounds are applied to $\Gamma(Z \to \gamma a) \times {\rm Br}(a \to \gamma\nu\bar\nu)$. 
\item
The LHC tri-boson searches give the best sensitivities on the ALP models in larger mass regions, especially above the $Z$ threshold.
In most of the ALP mass regions, the leading constraint is obtained by $pp \to \gamma^*, Z^* \to \gamma a$, $a \to Z\gamma \to \nu\bar\nu\gamma$.\footnote{Although the same final state is yielded by $pp \to aZ \to a\nu\bar\nu$ with $a \to \gamma\gamma$, the process is unlikely to contribute because the analysis requires a di-photon invariant mass larger than the ALP mass.}
Hence, the bounds are imposed on $\sigma(pp \to a\gamma) \times {\rm Br}(a \to Z\gamma)$. 
The results are obtained up to $m_a = 500\GeV$ in the literature.

The same process but with $Z \to \mu\bar\mu$ provides the best limit around $m_a = 110\GeV$.
Also, for general choices of $(c_{BB}, c_{WW})$ the constraints from $pp \to W^* \to W a$, $a \to W^+W^-$ may become significant especially around $m_a = 300\GeV$, though they should be compared to the constraints with $g_{a\gamma\gamma} \neq 0$.
\end{itemize}
These results are based on the assumption that the ALPs are produced directly, \ie, on-shell. 
In contrast, the reference~\cite{Bonilla:2022pxu} has studied non-resonant productions of the ALPs at the LHC. 
Here, they analyzed off-shell ALP contributions to vector-boson scattering processes such as $V_1V_2 \to a^* \to V_3V_4$ or t-channel ALP exchange, where $V_i$ denotes the gauge boson.
The constraints are derived for various $c_{BB}$, $c_{WW}$, and $m_a$.
Note that $m_{a}\lesssim 100~{\rm GeV}$ is assumed in the literature, otherwise on-shell ALP contributions cannot be neglected generally.

\section{Results} \label{sec: result}

In this section, we show the EWPT results of the ALPs and compare them with the experimental constraints.
Among the ALP contributions, the $Z$-boson decay into $a + \gamma$ affects the probability likelihood only when the ALP is lighter than the $Z$ boson. 
Hence, we will focus on the case of $m_{a} \ll m_{Z}$ in Sec.~\ref{sec: result_light_ALP}.
We will also show how much the overlooked contributions affect the EWPOs by comparing the results with those based on the analyses in the previous studies.

In Sec.~\ref{sec: result_heavier_ALP}, we will study the case when the ALP is not much lighter than the $Z$ boson. 
In Sec.~\ref{sec: ewpt}, we have provided the formulae of the ALP contributions that can be applied to $m_{a} \gtrsim m_{Z}$.
In this case, the ALPs are free from the flavor constraints but are subject to the collider ones, particularly from the LHC. 
We will compare the EWPT results with those bounds.

In Sec.~\ref{sec: result_W_mass}, we will discuss the goodness of fit by focusing on the $W$ mass.
In particular, the recent CDF result of the $W$ mass measurement is inconsistent with the SM prediction. 
It will be shown that the tension may be solved in the ALP model if the ALP is heavier than $500\GeV$, and thus, the goodness of fit is improved against the SM case.

\subsection{Light ALP case} \label{sec: result_light_ALP}

\begin{figure}[t]
    \begin{minipage}{0.5\linewidth}
        \centering
        \includegraphics[scale=0.35]{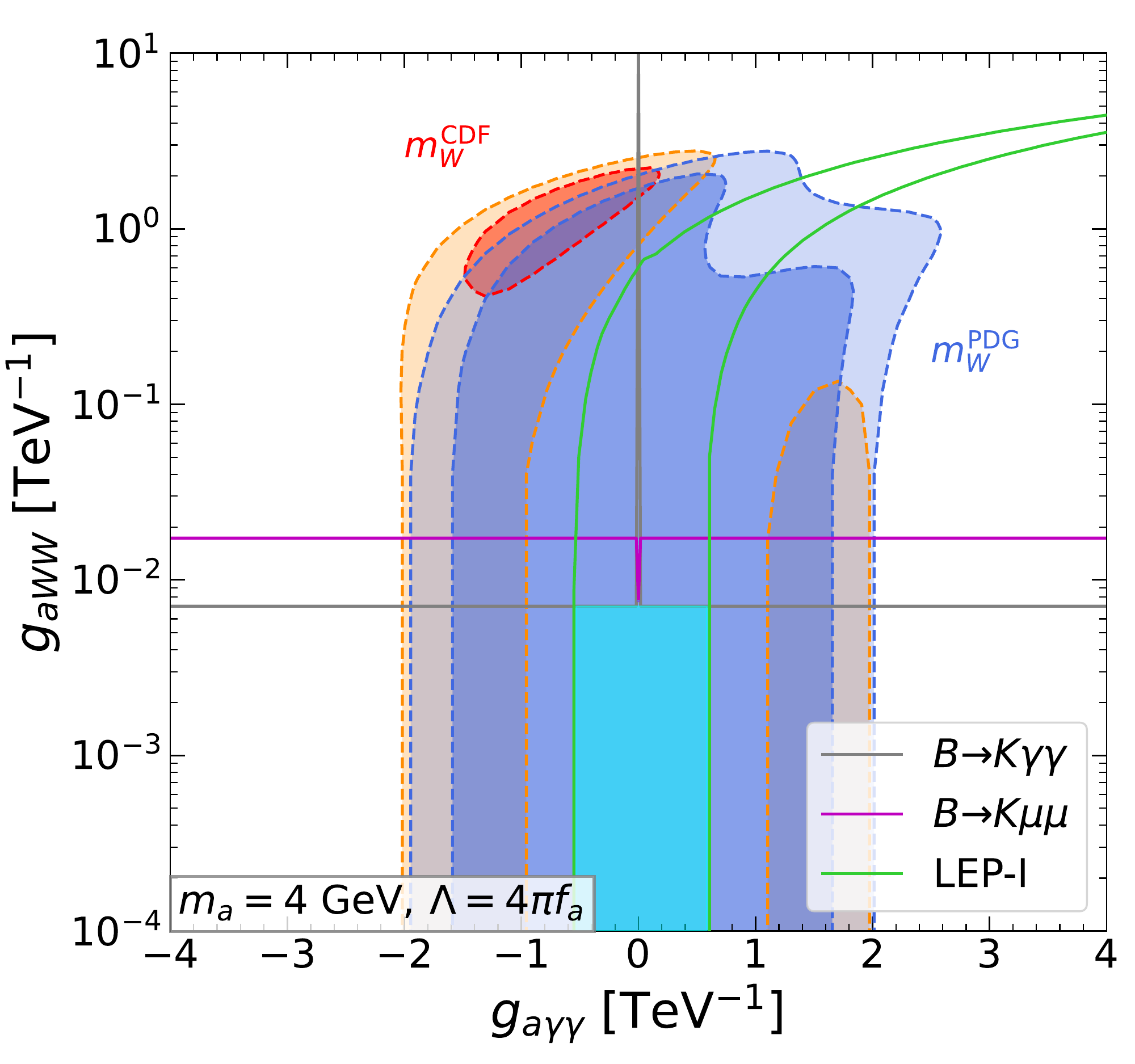}
    \end{minipage}
    \begin{minipage}{0.5\linewidth}
        \centering
        \includegraphics[scale=0.35]{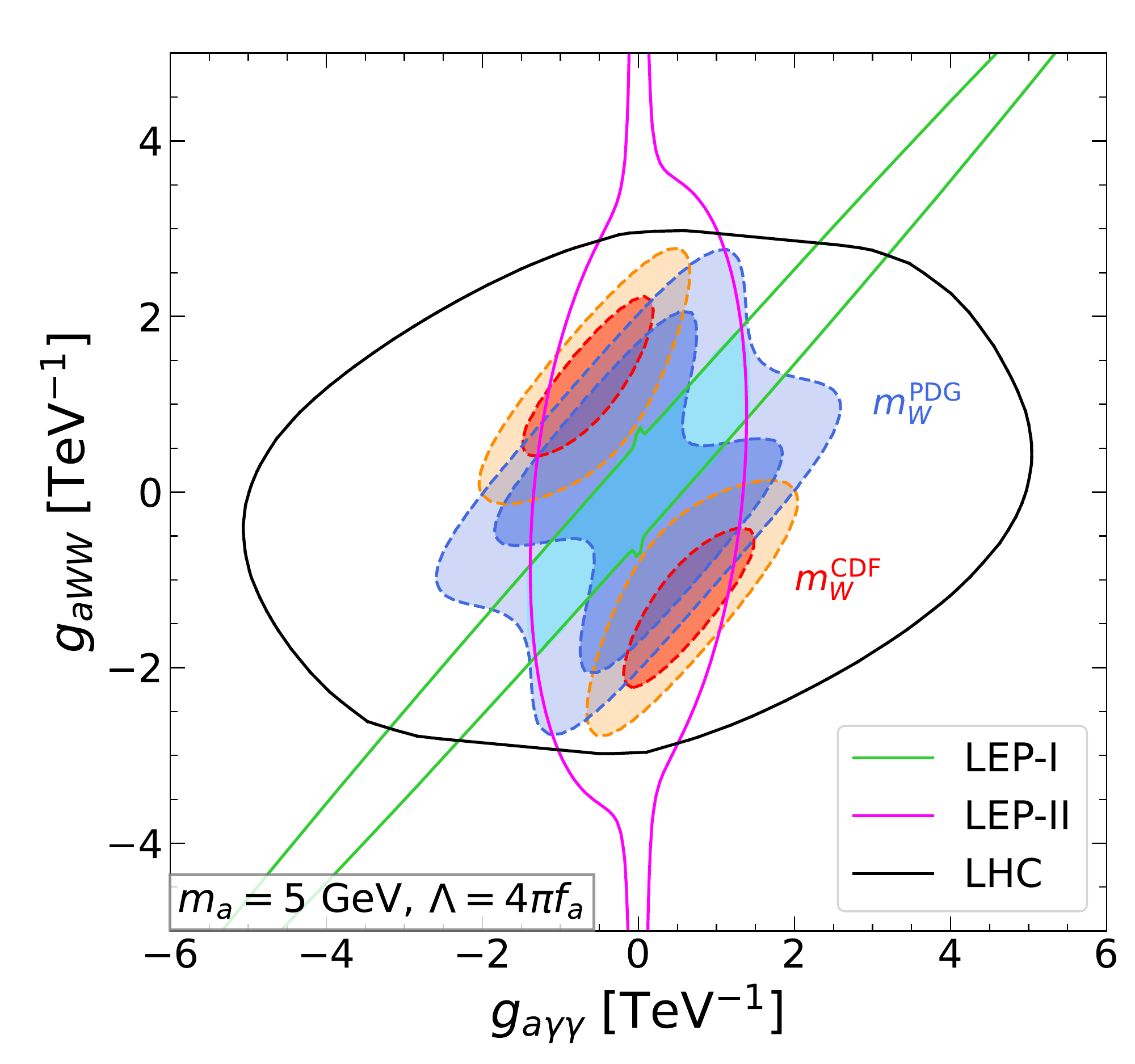}
    \end{minipage}
    \caption{EWPO probability distribution on the $(g_{a\gamma\gamma}, g_{aWW})$ plane in units of ${\rm TeV}^{-1}$.
    For the red (blue) colored regions, $m_{W}^{\rm CDF}$ $(m_{W}^{\rm PDG})$ is adopted as the experimental value of the $W$ mass. 
    The darker- (lighter-) colored regions enclosed by the dashed lines represent the 68\% (95\%) level.
    The ALP mass is $4\GeV$ (left) and $5\GeV$ (right). 
    Also, $f_{a}=1\TeV$ and $\Lambda=4\pi f_{a}$ are taken.
    The other colored solid lines denote the experimental constraints. 
    The regions including $(g_{a\gamma\gamma}, g_{aWW}) = (0,0)$ are allowed. 
    In the cyan region, all the constraints are satisfied. }
    \label{fig: EWPT_2D}
\end{figure}

Let us first consider the case when the ALP is much lighter than the $Z$ boson.
In this case, the ALP contributions to the EWPOs are insensitive to the ALP mass, as seen from Eqs.~\eqref{eq: delta_alpha_limit}--\eqref{eq: delta_W_limit}.
In Fig.~\ref{fig: EWPT_2D}, the EWPT results are shown for $f_a = 1\TeV$ and $\Lambda = 4\pi f_a$.
The probability likelihoods are calculated by globally fitting the ALP model to the EWPOs on the $(g_{a\gamma\gamma}, g_{aWW})$ plane. 
The red (blue) colored region is obtained by adopting $m_{W}^{\rm CDF}$ $(m_{W}^{\rm PDG})$ for the experimental data.
The probability distributions are normalized on the coupling parameter plane. 
The darker- (lighter-) colored regions enclosed by the dashed lines correspond to the 68\% (95\%) level.

On the left panel, the ALP mass is set as $m_{a} = 4\GeV$. 
The flavor constraints impose very severe limits on $|g_{aWW}|$.
The vertical axis is shown on a logarithmic scale to represent its tightness. 
In particular, the region above the gray line is excluded by $B^{+}\to K^{+}\gamma\gamma$.
Although this decay is suppressed around $g_{a\gamma\gamma}=0$, such a parameter region is instead constrained by $B^{+}\to K^{+}\mu\mu$, where the effective coupling $g_{a\mu\mu}^{\rm eff}$ is as large as or larger than $g_{a\gamma\gamma}$.
Then, the region above the magenta line is excluded. 
On the other hand, $g_{a\gamma\gamma}$ is limited by the collider constraints.
In particular, $e^{+}e^{-}\to\gamma\gamma(\gamma)$ at LEP-I provides the best sensitivity except for $g_{a\gamma\gamma} \simeq 0$, and $e^{+}e^{-}\to\gamma jj$ on the $Z$ pole (LEP-I) does around $g_{a\gamma\gamma} = 0$. 
These collider bounds are satisfied in the region between the green lines.

\begin{figure}[t]
    \begin{minipage}{0.5\linewidth}
        \centering
        \includegraphics[scale=0.35]{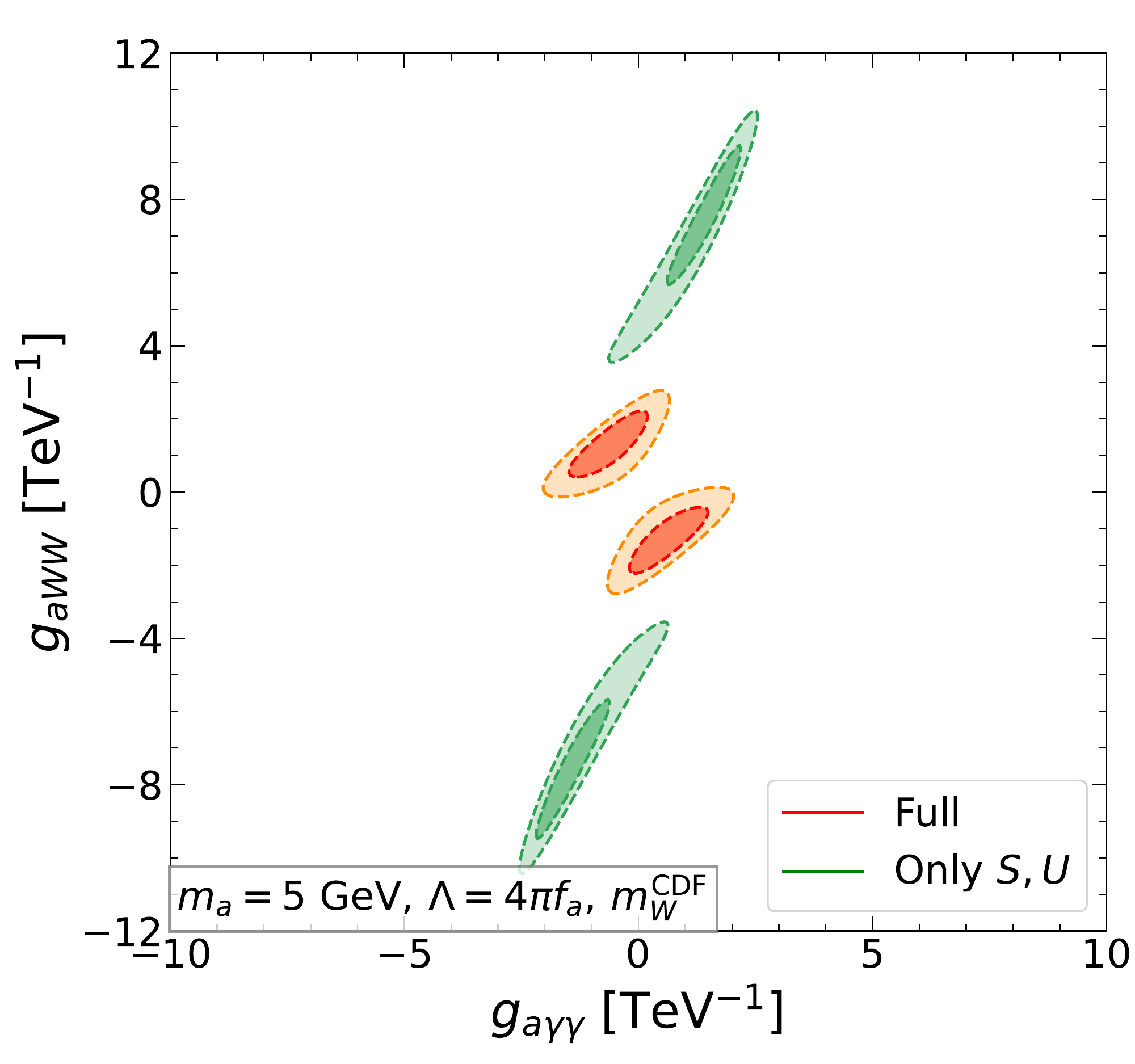}
    \end{minipage}
    \begin{minipage}{0.5\linewidth}
        \centering
        \includegraphics[scale=0.35]{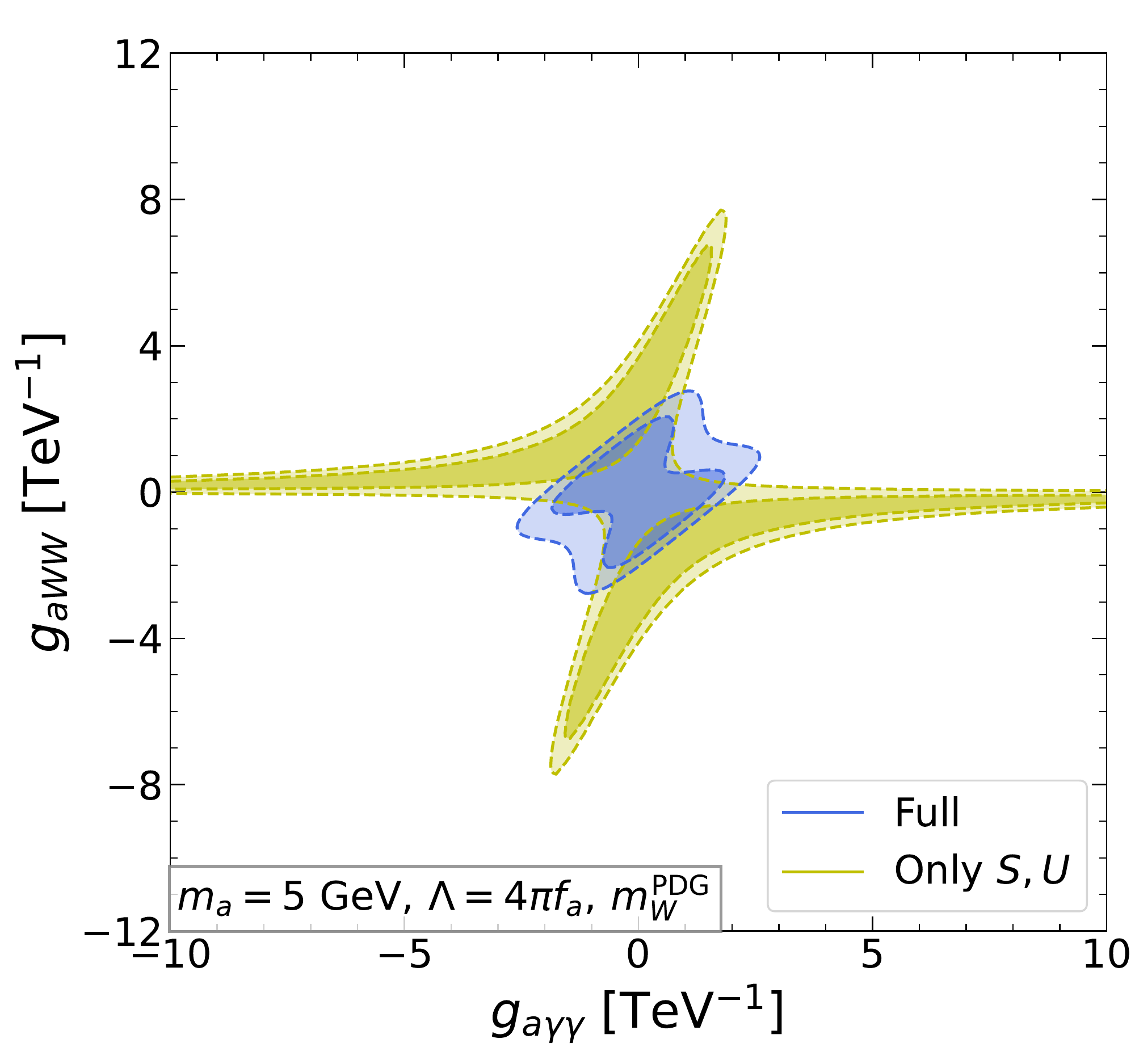}
    \end{minipage}
    \caption{EWPO probability distribution on $(g_{a\gamma\gamma}, g_{aWW})$.
    Here, $m_{a} = 5~{\rm GeV}$, $f_{a}=1\TeV$, and $\Lambda=4\pi f_{a}$ are taken.
    The red and blue regions are the same as Fig.~\ref{fig: EWPT_2D}. 
    On the left (right) panel, the green (yellow) regions are the EWPT result only via $S$ and $U$, \ie, ignoring the ALP contributions via $\Delta\alpha$, $\Delta_{Z}$, $\Delta_{W}$ and $\Gamma_{a\gamma}$.
    The darker (lighter) regions with the dashed lines correspond to the 68\% (95\%) level. }
    \label{fig: EWPT_2D_comp}
\end{figure}

On the right panel, the ALP mass is set as $m_{a} = 5\GeV$.
The EWPT results are almost the same as those on the left panel: Note that the vertical axis is shown on a linear (logarithmic) scale on the right (left).
For this ALP mass, the flavor constraints are absent because the heavy-meson decay into the ALP is kinematically blocked, and the collider bounds become relevant.
The green lines correspond to the constraints from the LEP-I data, as in the case with $m_{a}=4\GeV$.
Since this constraint is governed by $g_{aZ\gamma}$, the region around $g_{aZ\gamma} = 0$ is allowed.
The scattering $e^{+}e^{-}\to\gamma\gamma(\gamma)$ at LEP-II also gives an independent constraint, and the region inside the magenta lines is allowed. 
In particular, this constraint works even for $g_{aZ\gamma} \simeq 0$, because a photon-exchange diagram contributes to the scattering.
Although the LHC constraint by the non-resonant searches for the ALPs is shown by the black line, where the region inside the line is allowed, the result is weaker than those from LEP~I and II.

In both plots, the regions allowed by all the experimental constraints are filled by the cyan.\footnote{
In addition to the constraints discussed in Fig.~\ref{fig: EWPT_2D}, the Higgs boson decaying into a pair of ALPs, $h \to aa$, may limit the model. 
We have checked that it is weaker than those from the flavor and collider experiments.
}
It is found that the EWPT result becomes consistent with the constraints at the 68\% level only when the PDG value is adopted for the $W$ mass.
Besides, although the region around $g_{aZ\gamma}=0$ is relatively favored by the constraints, the probability likelihood of the EWPOs is not improved well.
Therefore, we conclude that the recent CDF result of $m_{W}$ cannot be explained by the ALP as long as the mass is $m_{a} \ll m_{Z}$. 

Let us compare our results with those based on the analyses in the previous studies~\cite{Bauer:2017ris, Bauer:2018uxu, Yuan:2022cpw}.
They considered the ALP much lighter than the $Z$ boson, which is the same as the setup in this subsection.
However, as mentioned in Sec.~\ref{sec: ewpt}, the contributions via $\Delta\alpha$, $\Delta_{Z}$, $\Delta_{W}$ and $\Gamma_{a\gamma}$ have been ignored, \ie, the ALP contributions have been supposed to arise only via $S$ and $U$. 
In Fig.~\ref{fig: EWPT_2D_comp}, we show the EWPT results with and without including $\Delta\alpha$, $\Delta_{Z}$, $\Delta_{W}$ and $\Gamma_{a\gamma}$. 
Here, $m_{a}=5~{\rm GeV}$, $f_a=1\TeV$ and $\Lambda=4\pi f_{a}$ are set.
The former result is the same as those on the right panel in Fig~\ref{fig: EWPT_2D}, while the latter corresponds to the setup in the previous studies.\footnote{
The ``only $S, U$'' regions in the figure are slightly different from those in the previous papers because the global-fit procedures are not the same.
For example, $T=0$ is not fixed in Ref.~\cite{Yuan:2022cpw} when the parameter regions of $S$ and $U$ are determined.
In contrast, this condition is satisfied in our analysis because the ALP does not contribute to $T$. 
}
On the left (right) panel, $m_{W}^{\rm CDF}$ $(m_{W}^{\rm PDG})$ is adopted as the experimental value of the $W$ mass. 
In both cases, it is obvious that the EWPT results are modified drastically by the new contributions.
In particular, the impact of $Z\to a\gamma$ is strong, because it proceeds at the tree level, and thus, worsens the global fit significantly in the region with $g_{aZ\gamma} \neq 0$.
In addition, the effects via $\Delta\alpha$, $\Delta_{Z}$, and $\Delta_{W}$ are comparable to those via $S$ and $U$.
Therefore, it is found that the analyses based only on $S$ and $U$ are not valid.
The ALP contributions via $\Delta\alpha$, $\Delta_{Z}$, $\Delta_{W}$ and $\Gamma_{a\gamma}$ must be taken into account in the EWPO analysis.

\subsection{Heavier ALP case} \label{sec: result_heavier_ALP}

\begin{figure}[t]
    \begin{minipage}{0.5\linewidth}
        \centering
        \includegraphics[scale=0.35]{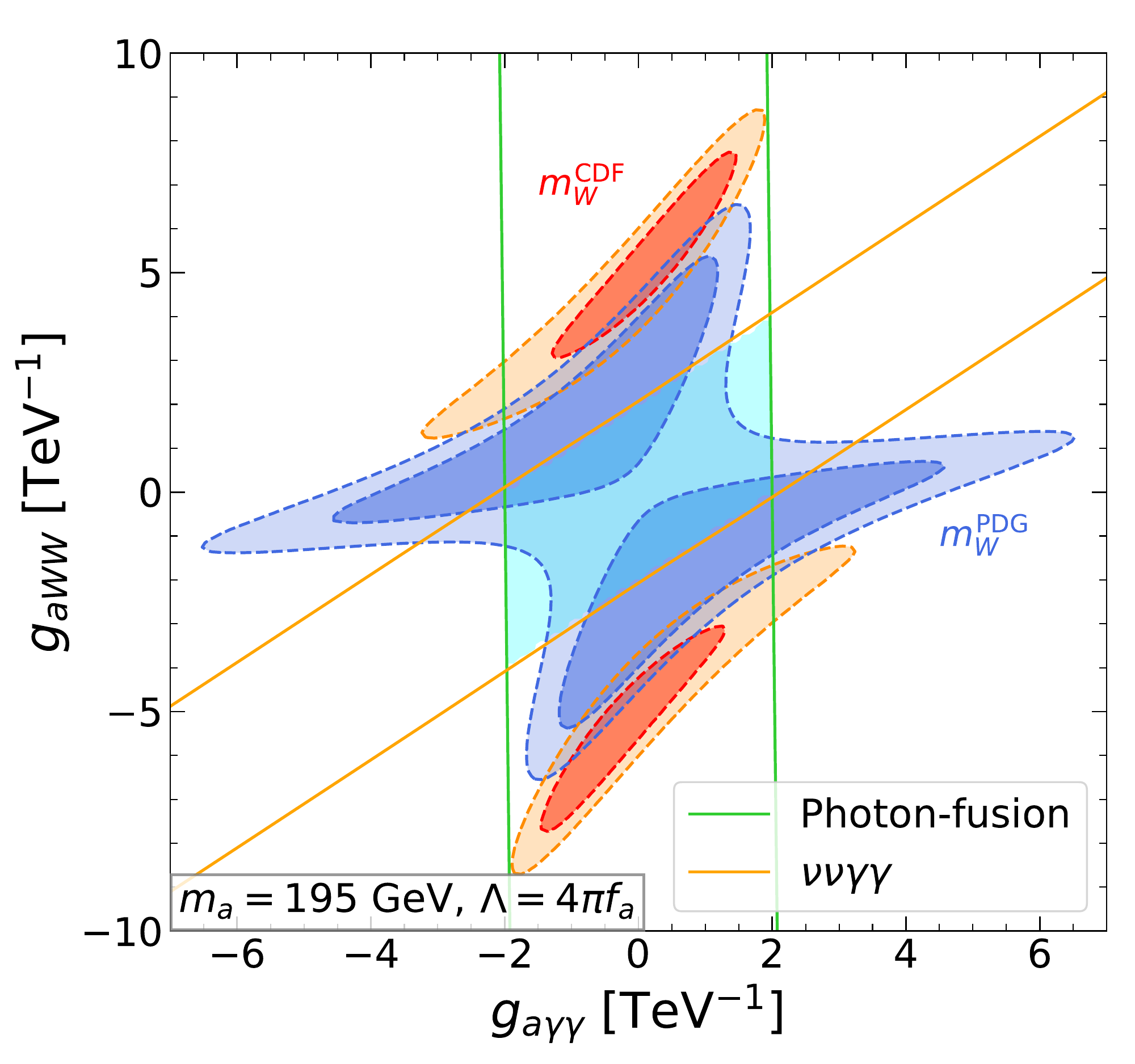}
    \end{minipage}
    \begin{minipage}{0.5\linewidth}
        \centering
        \includegraphics[scale=0.35]{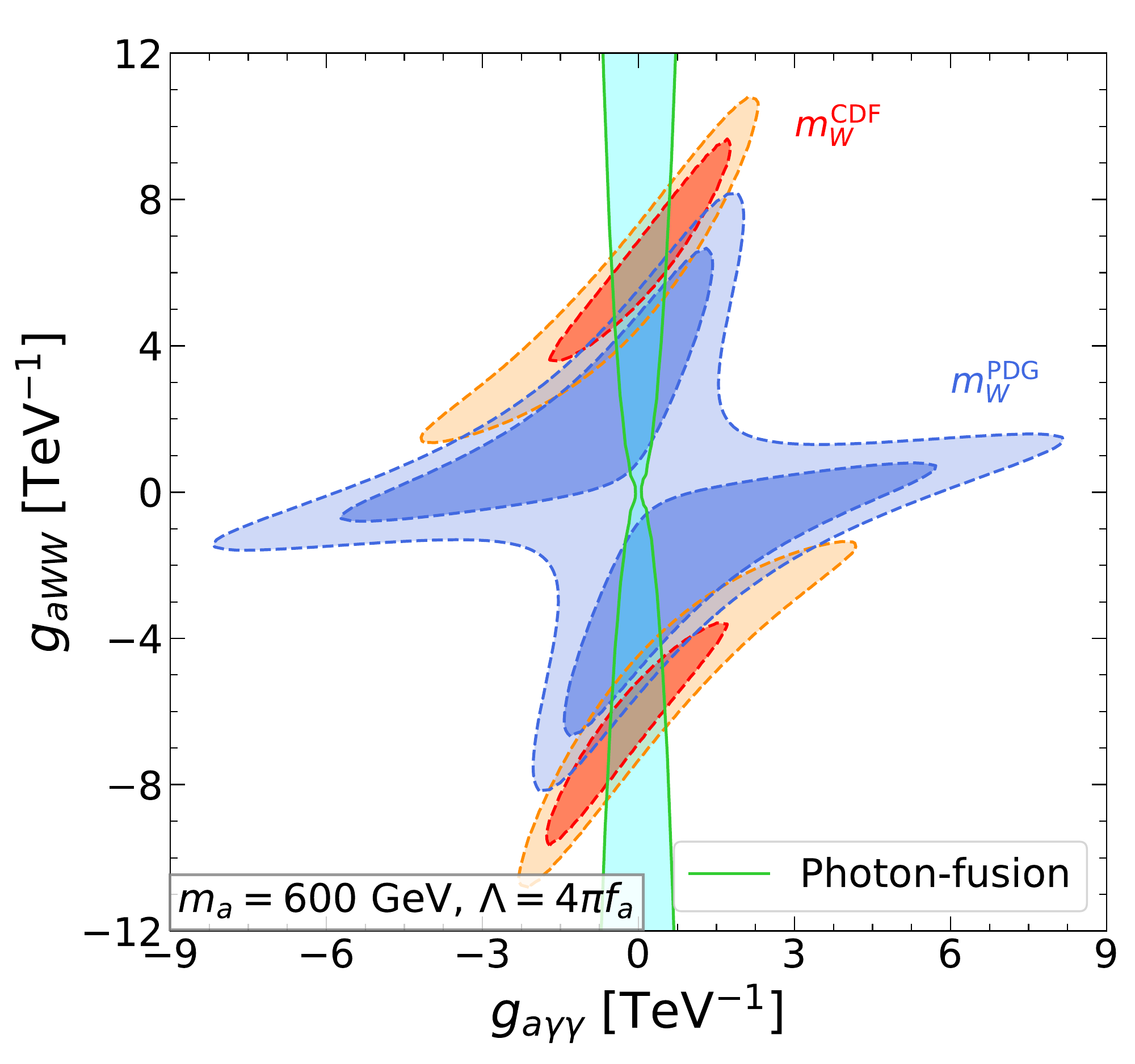}
    \end{minipage}
    \caption{Same as Fig.~\ref{fig: EWPT_2D} but $m_{a} = 195\GeV$ (left) and $600\GeV$ (right). 
    The collider bounds are shown by the orange and green solid lines, and both of them are satisfied in the cyan region. }
    \label{fig: EWPT_2D_ma_gtr_mZ}
\end{figure}

As shown in the previous subsection, the EWPT results for $m_{a} \ll m_{Z}$ are tightly limited by the flavor and collider constraints.
In this subsection, we consider the case of $m_{a} \gtrsim m_{Z}$. 
In Fig.~\ref{fig: EWPT_2D_ma_gtr_mZ}, we show the EWPT results on the $(g_{a\gamma\gamma}, g_{aWW})$ plane.
Here, we take $m_{a} = 195\GeV$ and $600\GeV$ on the left and right panels, respectively.
Also, $\Lambda=4\pi f_{a}$ with $f_{a}=1\TeV$ is set for both masses. 

The collider constraints depend on the ALP mass.
For $m_{a}=195~{\rm GeV}$, the photon scattering via the off-shell ALP exchange, $\gamma\gamma\to a^{*}\to \gamma\gamma$, gives the best constraint on $g_{a\gamma\gamma}$. 
The region between the green lines is allowed by this bound.
On the other hand, the constraint from $pp\to Z^{*}\to a\gamma$ with $a\to Z\gamma\to\nu\nu\gamma$ is shown by the orange lines, where the region between the lines is allowed. 
Note that the production cross section becomes smaller as $g_{aZ\gamma}$ is suppressed. 
The region that satisfies both constraints is shown by the cyan. 
It is found that the EWPT result is consistent with them at the 68\% level if the PDG value is adopted for the $W$ mass, while there is no overlap for the CDF value.

As mentioned in Sec.~\ref{sec: constraint}, many of the collider bounds can be avoided if the ALP is heavier than $500\GeV$. 
From the right panel, where the mass is $m_{a}=600~{\rm GeV}$, only the search for the on-shell ALP production via the photon fusion, $\gamma\gamma\to a\to \gamma\gamma$, gives a constraint. 
This is displayed by the green lines, and the region between them is allowed (filled by the cyan). 
It is seen that the ALP coupling is favored to satisfy $g_{a\gamma\gamma} \simeq 0$, and the constraint becomes weaker as $g_{aWW}$ grows because the branching ratio of $a\to\gamma\gamma$ decreases.
We found that the EWPT result can be consistent with the constraints at the 68\% level even for $m_{W}^{\rm CDF}$ as well as $m_{W}^{\rm PDG}$. 
The $W$ mass will be discussed more in Sec.~\ref{sec: result_W_mass}. 

\begin{figure}[t]
    \centering
    \includegraphics[scale=0.5]{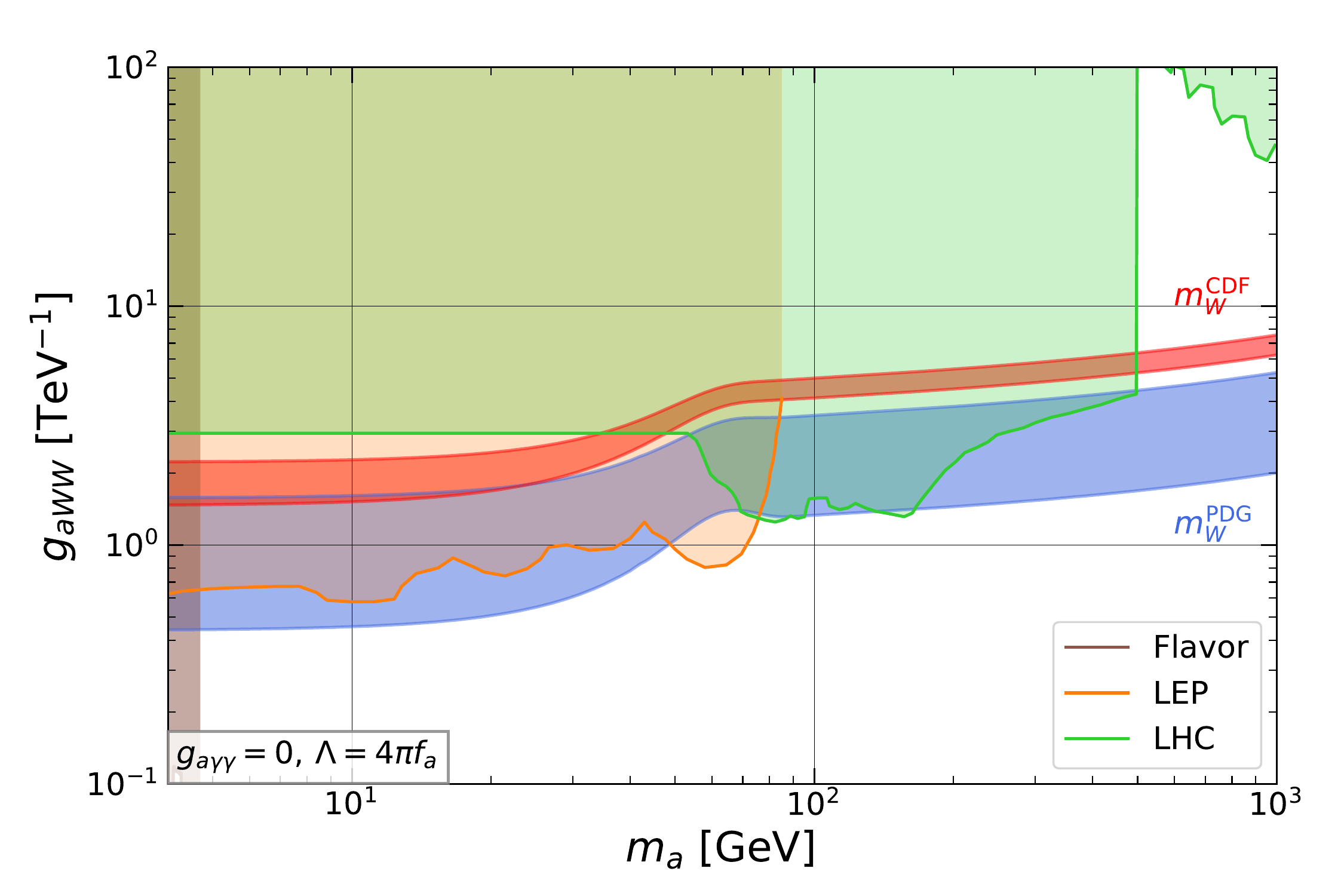}
    \caption{EWPO probability distribution for various $m_{a}$. 
    Here, $g_{a\gamma\gamma}=0$, $f_{a}=1\TeV$, and $\Lambda=4\pi f_{a}$ are taken.
    The distribution function is normalized on the $g_{aWW}$ parameter region for each $m_{a}$.
    The red (blue) colored region corresponds to the 68\% level for $m_{W}^{\rm CDF}$ $(m_{W}^{\rm PDG})$.
    The regions excluded by the flavor, LEP, and LHC constraints are filled by brown, orange, and green, respectively.}
    \label{fig: EWPT_1D_Lam4pi}
\end{figure}

As the ALP coupling to di-photon is favored to be suppressed, let us focus on the case with $g_{a\gamma\gamma} = 0$. 
In Fig.~\ref{fig: EWPT_1D_Lam4pi}, we show the EWPT result for various $m_{a}$ with $g_{a\gamma\gamma} = 0$ and compare them with the experimental constraints. 
Here, we take $\Lambda=4\pi f_{a}$ with $f_{a}=1\TeV$. 
The probability distribution is normalized on the $g_{aWW}$ parameter space for each $m_{a}$, and the 68\% probability range is shown. 
The result for 95\% probability range is given in Appendix~\ref{app: gaAA0_95}.
The red (blue) colored band is the result obtained for $m_{W}^{\rm CDF}$ $(m_{W}^{\rm PDG})$. 
The regions excluded by the flavor, LEP, and LHC are shown by the brown, orange, and green regions, respectively.
It is found that the EWPT result for $m_{W}^{\rm PDG}$ can be consistent with the constraints at the 68\% level except for $50 \lesssim m_{a} \lesssim 160\GeV$, though a part of the parameter regions is excluded for $m_{a} \lesssim 50\GeV$ (see also the discussion in Sec.~\ref{sec: result_W_mass}).
In contrast, the result for $m_{W}^{\rm CDF}$ is already excluded unless the ALP is heavier than $500\GeV$.

\subsection{Goodness of fit and $W$-boson mass} \label{sec: result_W_mass}

In Secs.~\ref{sec: result_light_ALP} and \ref{sec: result_heavier_ALP}, we showed the EWPT results, where the likelihood functions were studied by performing the global fit to the EWPOs.
It should be stressed that the probability distribution was normalized on the ALP-coupling plane, \ie, the total probability is equal to unity on the plane, for each ALP mass.
Therefore, the parameter region corresponding to ``68\% probability'' does not always mean that {\it all} theoretical values are consistent with the experimental data but indicates that the fit inside the region is {\it relatively} better than the outside. 
For instance, the tension between the theoretical and CDF values of the $W$ mass is not always solved even in that region. 
In this subsection, let us discuss how well the ALP contributions improve the global fit, particularly paying attention to the case when $m_{W}^{\rm CDF}$ is adopted to the likelihood.

\begin{figure}[t]
    \centering
    \includegraphics[scale=0.7]{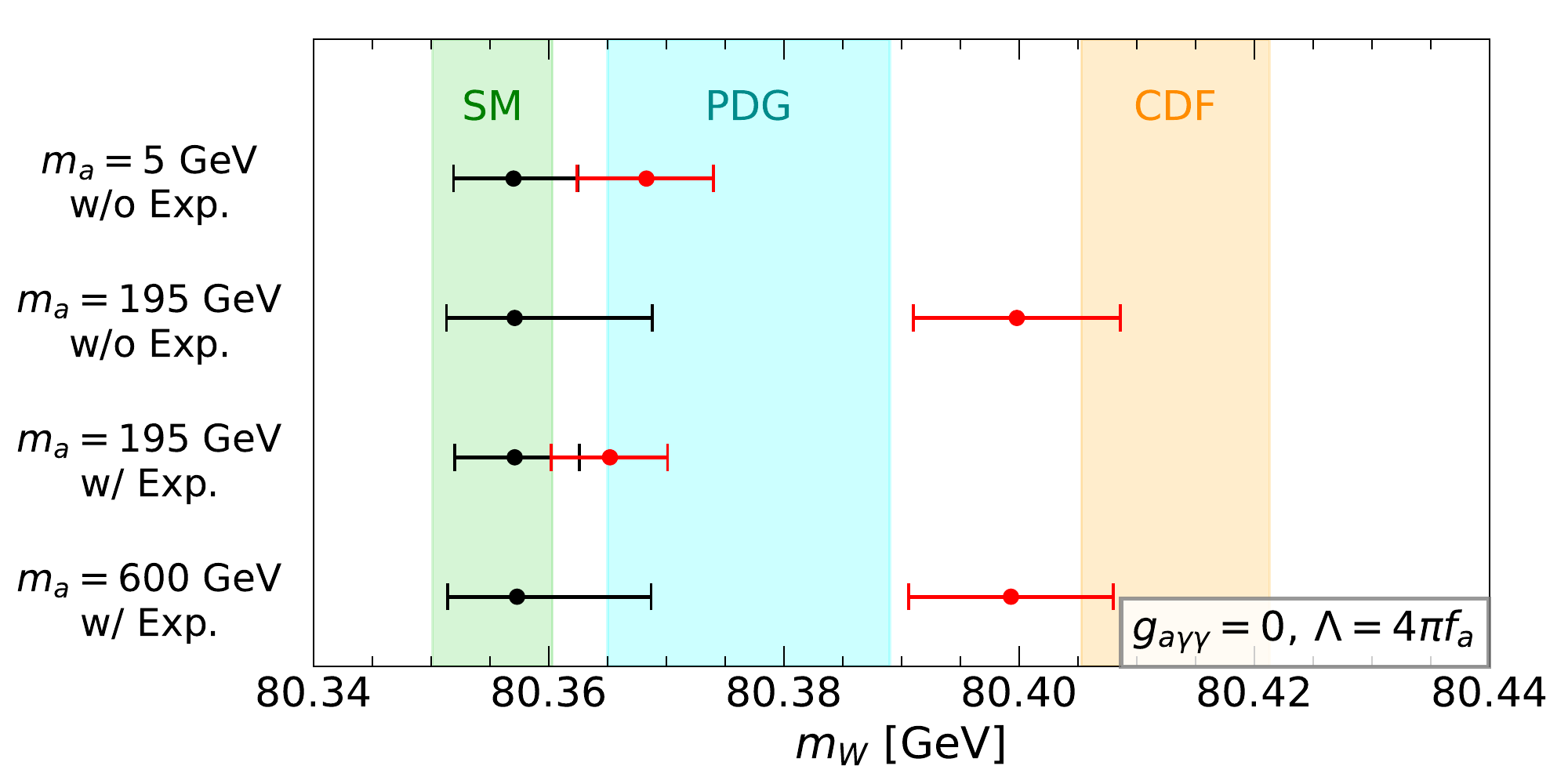}
    \caption{ALP predictions of $m_{W}$.
    Here, $g_{a\gamma\gamma}=0$, $f_{a}=1\TeV$, and $\Lambda=4\pi f_{a}$ are taken.
    The red (black) bars represent the results obtained with/without including $m_{W}$ in the global fit.
    The orange and cyan bands correspond to $m_{W}^{\rm CDF}$ and $m_{W}^{\rm PDG}$ with the $1\sigma$ uncertainty, while the green band is the SM prediction with the $1\sigma$ error. 
    The ALP coupling is restricted by the experimental constraints during the fit analysis for the results with the label ``w/ Exp.''}
    \label{fig: MwValue}
\end{figure}

Since the largest tension arises in the $W$ mass, the goodness of fit is governed by $m_{W}$ among the EWPOs listed in Table~\ref{tab:EWPO}. 
In Fig.~\ref{fig: MwValue}, we show the theoretical values of $m_{W}$ for various $m_{a}$.
Here, $g_{a\gamma\gamma} = 0$, $f_{a}=1\TeV$ and $\Lambda=4\pi f_{a}$ are set.
The results are obtained by performing the global fit to the EWPOs.
In particular, the indirect prediction of $m_{W}$ is displayed by the black bar, where $m_{W}$ is not included in the fit observables. 
On the other hand, the red bar represents the theoretical value for which $m_{W}$ is included in the likelihood. 
We derive the probability distributions as a function of $m_{W}$. 
Then, the central value, which is denoted by the dot on the bar, is determined such that the likelihood becomes maximum, while the 68\% uncertainties are shown by the bars. 
The results are compared with the SM prediction (green), PDG (cyan), and CDF-averaged (orange) values, where the $1\sigma$ range is shown. 

For $m_{a}=5\GeV$, the black and red bars are obtained without taking the experimental constraints into account. 
Even though the constraints are ignored, the theoretical values cannot be consistent with $m_{W}^{\rm CDF}$ because of the effects of $Z \to a\gamma$. 
Since the ALP contribution deteriorates the fit rapidly, the model cannot change the theoretical values of the EWPOs against the SM prediction, \ie, does not solve the $m_{W}$ tension. 
Although the result is shown for $m_{a}=5\GeV$, the same conclusion holds as long as $Z \to a\gamma$ is effective, \ie, for $m_{a} \lesssim 90\GeV$.
Similarly, the goodness of fit for the case with $m_{W}^{\rm PDG}$ is not improved so much against the SM case. 

If the ALP is heavier than the $Z$ boson, the decay of $Z \to a\gamma$ is kinematically blocked. 
We show the results for $m_{a}=195\GeV$ with and without restring the parameter space by the experimental constraints, which are shown as the label, ``w/o Exp'' and ``w/ Exp,'' respectively.
According to the former result, although the indirect prediction seems to be around the SM value and does not overlap with the $m_{W}^{\rm CDF}$ range, the uncertainty is asymmetric between the smaller and larger $m_{W}$ values. 
Namely, the probability distribution has a long tail toward larger $m_{W}$. 
Consequently, the theoretical prediction with including $m_{W}$ in the fit (the red bar) becomes consistent with $m_{W}^{\rm CDF}$ at the 68\% level.

Once the experimental constraints are taken into account, the parameter space is restricted tightly for $m_{a}=195\GeV$.
Then, the ALP contribution cannot make $m_{W}$ shift from the SM prediction sufficiently, as shown by the result for ``w/ Exp.''
It is understood from Fig.~\ref{fig: EWPT_1D_Lam4pi} that the same conclusion holds as long as the collider bounds are tight. 

The collider constraints as well as those from the flavor are relaxed if the ALP is heavier than  $500\GeV$. 
In the table, we show the theoretical values for $m_{a} = 600\GeV$.
As shown in Fig.~\ref{fig: EWPT_2D_ma_gtr_mZ}, the model is free from the experimental constraints for $g_{a\gamma\gamma} = 0$. 
Similar to the case for $m_{a}=195\GeV$ without including the experimental constraints, it is found that the ALP contributions can solve the tension between the SM and CDF-averaged values of the $W$ mass. 

In summary, although the quality of the global fit to the EWPOs can be improved effectively against the SM case by the ALP with a mass larger than $m_{Z}$, the model is constrained tightly by the collider measurements for $m_{a}<500\GeV$. 
Thus, the tension between the SM and CDF-averaged values of the $W$ mass can be solved only when the ALP is heavier than $500\GeV$.

In Fig.~\ref{fig: MwValue}, we have focused on the ALP model with $m_{W}^{\rm CDF}$ for the experimental value of the $W$ mass. 
Here, let us comment on the case when $m_{W}^{\rm PDG}$ is adopted. 
Similar to the above case, the probability likelihood is less affected by the ALP as long as $Z \to a\gamma$ and/or the experimental bounds are severe. 
The goodness of fit is improved effectively compared to the SM case if the ALP is heavier than about $160\GeV$ under the collider constraints (see also Fig.~\ref{fig: EWPT_1D_Lam4pi}).

\begin{figure}[t]
    \centering
    \includegraphics[scale=0.5]{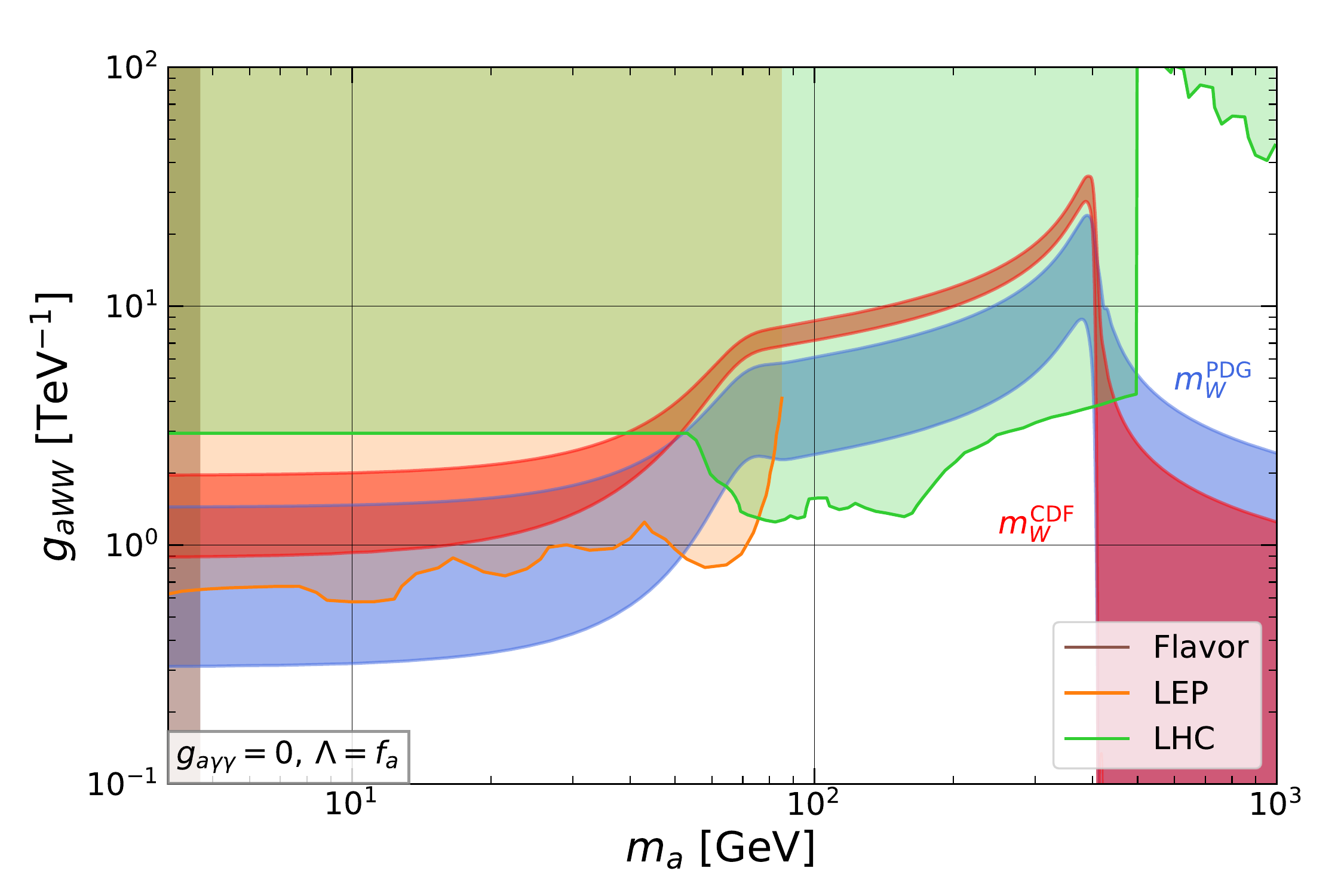}
    \caption{Same as Fig.~\ref{fig: EWPT_1D_Lam4pi} but $\Lambda=f_{a}$.}
    \label{fig: EWPT_1D_Lam1}
\end{figure}

Before closing this section, let us comment on the cutoff dependence of the results.
The cutoff appears in the loop functions and should be determined by a UV theory of the ALP model.
In Fig.~\ref{fig: EWPT_1D_Lam1}, we perform the same analysis as Fig.~\ref{fig: EWPT_1D_Lam4pi} but with $\Lambda=f_{a}$.
It is noticed that the EWPT regions drop rapidly above $m_{a}=400\GeV$.
This is because the ALP contributions especially to the $W$ mass flip their signs. 
As a result, the goodness of fit is not improved well by the ALP contributions.
Although the $m_{W}$ tension seems to be solved for $90 \lesssim m_{a} \lesssim 400\GeV$, all the parameter regions are excluded by the experimental constraints. 
Therefore, if the recent CDF result is confirmed in future experiments, a larger cutoff scale is favored.

\section{Conclusions and discussion} \label{sec: conclusion}

We have studied the EWPOs in the ALP model. 
The ALP is assumed to couple with the SM ${\rm SU(2)}_L$ and the ${\rm U(1)}_Y$ gauge bosons. 
We have provided the formulae of the ALP contributions to the EWPOs valid for any ALP mass and shown that those to $\Delta\alpha$, $\Delta_{Z}$, and $\Delta_{W}$ as well as the oblique parameter $U$ can be comparable to $S$ ($T=0$ in the ALP model). 
Furthermore, the decay of $Z\to a\gamma$ contributes to the total width of the $Z$-boson decay significantly. 

We have analyzed the probability likelihood generated from the ALP contributions by performing the global fit to the EWPOs and compared the results with the flavor and collider constraints. 
Because of $Z\to a\gamma$, the ALP contributions to the EWPOs are suppressed as long as the ALP is lighter than the $Z$ boson.
Besides, since the experimental constraints are tight, the EWPT regions are widely excluded already if the ALP is not so heavy.
Also, even when the ALP is heavy, the parameter regions are limited as long as the ALP is coupled sizably with a pair of photons. 
Therefore, a heavy ALP with interactions satisfying $g_{a\gamma\gamma} \sim 0$ is preferred to improve the global fit compared to the SM case.

In the numerical analysis, we have studied both of the cases when the recent CDF result of the $W$ mass measurement is included and excluded in the experimental data. 
If the PDG value adopted, \ie, the CDF result is neglected in the $W$ mass average, the ALP  is favored to be heavier than about $160\GeV$ with satisfying $g_{a\gamma\gamma} \sim 0$. 
On the other hand, if the recent CDF result, which is inconsistent with the SM prediction, is included, it has been concluded that the ALP model can solve the tension only for $m_{a} > 500\GeV$ with $g_{a\gamma\gamma} \sim 0$. 

Let us comment on contributions from higher-dimensional operators that are not included in the ALP Lagrangian \eqref{eq: Lagrangian}. 
They are likely to be non-negligible when $m_{a}$ is comparable to $f_a$. 
The higher-dimensional operators should depend on a UV theory of the ALP model. 
The theory also determines the cutoff scale, which has been set as a model parameter in our analysis. 
Since we have set $f_{a}=1\TeV$, the results might be altered especially around $m_{a} \sim 1\TeV$.

We have argued that the EWPOs are sensitive to the ALP models. 
In particular, if the recent CDF result of the $W$ mass measurement would be confirmed in future experiments, the model could provide an attractive solution. 
Although the model parameters are restricted by the experimental constraints, the model setup for solving the tension has not been explored sufficiently by colliders such as the LHC. 
Therefore, further collider studies would be helpful to establish the scenario.

\section*{Acknowledgements}
This work is supported by the Japan Society for the Promotion of Science (JSPS) Grant-in-Aid for Scientific Research on Innovative Areas (No.~21H00086 [ME] and No.~22J01147 [MA]) and Scientific Research B (No.~21H01086 [ME]). 

\appendix
\section{Passarino-Veltman functions} \label{app: PV_funcs}
The Passarino-Veltman functions~\cite{Passarino:1978jh} are denoted explicitly as
\begin{align}
    A_{0}(m_{0})
    &=
    \int\frac{\overline{\dd^{d}k}}{i\pi^{2}}
        \frac{1}{k^{2}-m_{0}^{2}+i\epsilon}
    = m_{0}^{2}\qty(1-\ln{\frac{m_{0}^{2}}{\mu^{2}}}),
    \\
    B_{0}(p^{2}; m_{0},m_{1})
    &=
    \int\frac{\overline{\dd^{d}k}}{i\pi^{2}}
        \frac{1}{[k^{2}-m_{0}^{2}+i\epsilon][(k+p)^{2}-m_{1}^{2}+i\epsilon]}
    \notag \\ &=
    -\int_{0}^{1}\dd x \ln{\qty[\frac{-x(1-x)p^{2}+xm_{0}^{2}+(1-x)m_{1}^{2}}{\mu^{2}}]},
\end{align}
where $d=4-2\epsilon$ and $\overline{\dd^{d}k}=\Gamma(1-\epsilon)(\pi\mu^{2})^{\epsilon}\dd^{d}k$ in the $\overline{\rm MS}$ regularization~\cite{Hagiwara:1994pw}.
We take $\mu=\Lambda$ in this paper.
The derivative of the $B_{0}$ function with respect to $p^{2}$ is obtained as
\begin{align}
    B'_{0}(p^{2}; m_{0},m_{1})
    &=
    \int_{0}^{1}\dd x
        \frac{x(1-x)}{-p^{2}x(1-x)+xm_{0}^{2}+(1-x)m_{1}^{2}}.
\end{align}

\section{Three-body decay of ALP} \label{app: aToZstarA}

The decay width for $a\to Z^{*}\gamma\to f f \gamma$ is derived at the tree level as
\begin{align}
\Gamma(a \to f\bar f \gamma) &=
N_{c}^{f}\frac{g_{Z}^{2}g_{aZ\gamma}^{2}}{1536\pi^{3}m_{a}^{3}}
\qty[(g_{V,f})^{2}+(g_{A,f})^{2}]
\int_{0}^{m_{a}^{2}}\dd x\frac{x(m_{a}^{2}-x)^{3}}{(x-m_{Z}^{2})^{2}+m_{Z}^{2}\Gamma_{Z}^{2}},
\end{align}
where the mass of the fermions in the final state is neglected.
The integral is evaluated analytically as
\begin{align}
&\int_{0}^{m_{a}^{2}}\dd x\frac{x(m_{a}^{2}-x)^{3}}{(x-m_{Z}^{2})^{2}+m_{Z}^{2}\Gamma_{Z}^{2}}
=\frac{1}{6}\Bigg\{
-11m_{a}^{6}+30m_{a}^{4}m_{Z}^{2}-18m_{a}^{2}m_{Z}^{4}+6m_{a}^{2}m_{Z}^{2}\Gamma_{Z}^{2}
\notag \\ &\quad
+3[3m_{Z}^{2}(3m_{Z}^{2}-\Gamma_{Z}^{2})m_{a}^{2}-4m_{Z}^{4}(m_{Z}^{2}-\Gamma_{Z}^{2})+m_{a}^{6}-6m_{a}^{4}m_{Z}^{2}]
\notag \\ &\qquad
\times \ln{\qty[\frac{(m_{a}^{2}-m_{Z}^{2})^{2}+m_{Z}^{2}\Gamma_{Z}^{2}}{m_{Z}^{2}(m_{Z}^{2}+\Gamma_{Z}^{2})}]}
\notag \\ &\quad
-6\frac{m_{Z}}{\Gamma_{Z}}[\Gamma_{Z}^{4}m_{Z}^{2}-3\Gamma_{Z}^{2}(m_{a}^{4}-3m_{a}^{2}m_{Z}^{2}+2m_{Z}^{4})-(m_{a}^{2}-m_{Z}^{2})^{3}]
\notag \\ &\qquad
\times\qty[\arctan{\qty(\frac{m_{Z}}{\Gamma_{Z}})}+\arctan{\qty(\frac{m_{a}^{2}-m_{Z}^{2}}{m_{Z}\Gamma_{Z}})}]
\Bigg\}.
\end{align}

\section{EWPO probability distribution for 95\% level.} \label{app: gaAA0_95}

\begin{figure}[t]
    \centering
    \includegraphics[scale=0.5]{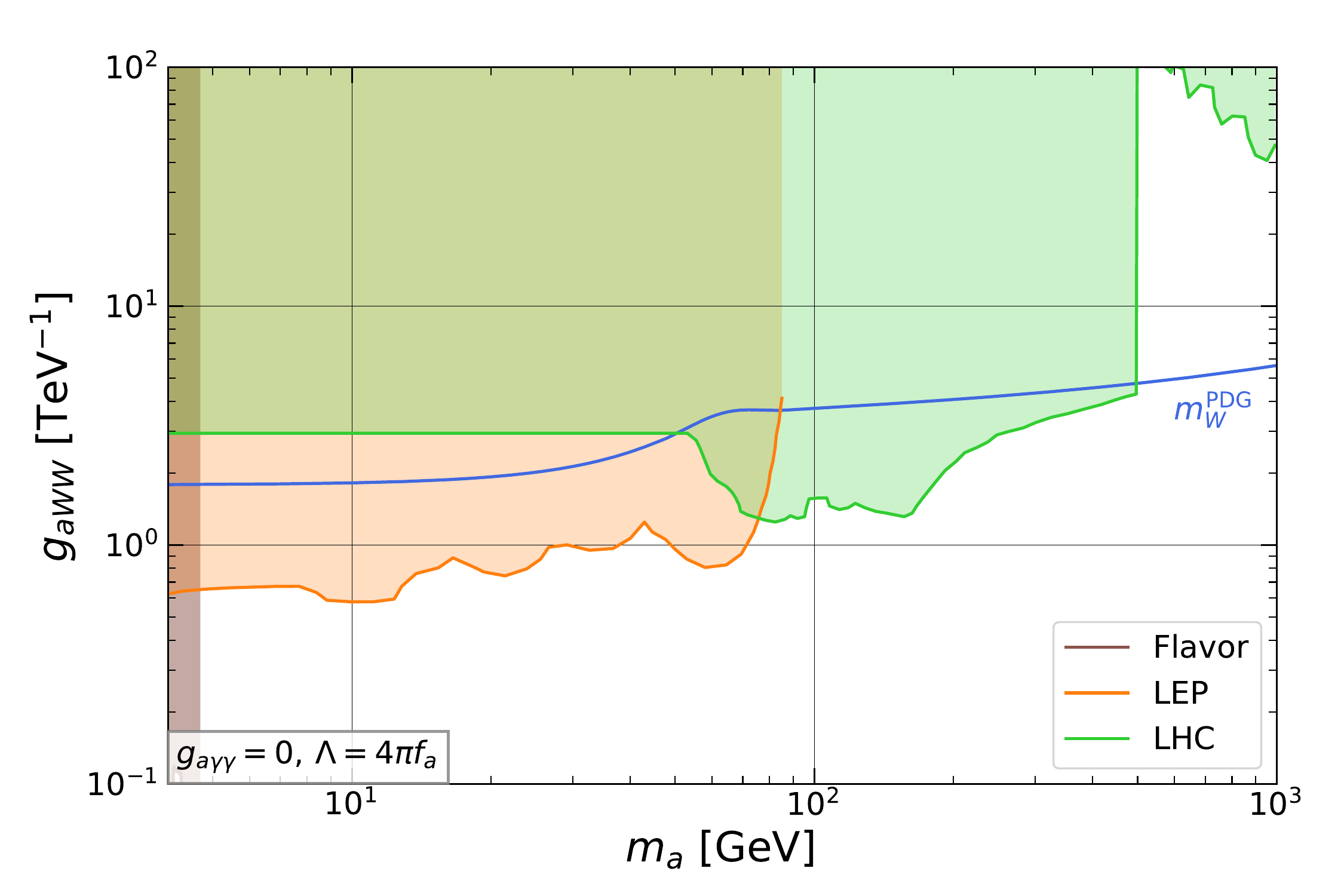}
    \caption{Upper bound from the EWPT on $g_{aWW}$ at the 95\% level (blue line). 
    Here, the experimental value of the $W$ mass is provided by $m_{W}^{\rm PDG}$.
    See Fig.~\ref{fig: EWPT_1D_Lam4pi} for the model setup and the constraints. }
    \label{fig: EWPT_1D_Lam4pi_95}
\end{figure}

In Fig.~\ref{fig: EWPT_1D_Lam4pi_95}, we show the upper bound from the EWPT on $g_{aWW}$ at the 95\% level (blue line).
Here, the experimental value of the $W$ mass is provided by $m_{W}^{\rm PDG}$.
Also, the model setup and the constraints are the same as Fig.~\ref{fig: EWPT_1D_Lam4pi}. 
Note that the EWPT result is consistent with the SM limit, \ie, $g_{aWW} = 0$, at the 95\% level.
It is found that the EWPT provides the most sensitive probe of the ALP for $m_{a} > 500\GeV$, and the model is disfavored if $g_{aWW}\gtrsim 4\text{--}6\TeV^{-1}$. 

\bibliographystyle{utphys28mod}
\bibliography{references}

\providecommand{\href}[2]{#2}\begingroup\raggedright\begin{thebibliography}{10}

\bibitem{Bauer:2017ris}
M.~Bauer, M.~Neubert, and A.~Thamm, {\em {Collider Probes of Axion-Like
  Particles},} \href{https://dx.doi.org/10.1007/JHEP12(2017)044}{JHEP
  {\bfseries 12} (2017) 044} {\ttfamily
  [\href{https://arxiv.org/abs/1708.00443}{arXiv:1708.00443}]}.

\bibitem{Peskin:1991sw}
M.~E.~Peskin and T.~Takeuchi, {\em {Estimation of oblique electroweak
  corrections},} \href{https://dx.doi.org/10.1103/PhysRevD.46.381}{Phys.\
  Rev.\  D {\bfseries 46} (1992) 381--409}.

\bibitem{CDF:2022hxs}
{\bfseries CDF} Collaboration, {\em {High-precision measurement of the $W$
  boson mass with the CDF II detector},}
  \href{https://dx.doi.org/10.1126/science.abk1781}{Science {\bfseries 376}
  (2022) 170--176}.

\bibitem{Yuan:2022cpw}
G.-W.~Yuan, L.~Zu, L.~Feng, Y.-F.~Cai, and Y.-Z.~Fan, {\em {Is the W-boson mass
  enhanced by the axion-like particle, dark photon, or chameleon dark energy?}}
  \href{https://dx.doi.org/10.1007/s11433-022-2011-8}{Sci.\  China Phys.\
  Mech.\  Astron.\  {\bfseries 65} (2022) 129512} {\ttfamily
  [\href{https://arxiv.org/abs/2204.04183}{arXiv:2204.04183}]}.

\bibitem{Baak:2014ora}
{\bfseries Gfitter Group} Collaboration, {\em {The global electroweak fit at
  NNLO and prospects for the LHC and ILC},}
  \href{https://dx.doi.org/10.1140/epjc/s10052-014-3046-5}{Eur.\  Phys.\  J.\
  C {\bfseries 74} (2014) 3046} {\ttfamily
  [\href{https://arxiv.org/abs/1407.3792}{arXiv:1407.3792}]}.

\bibitem{ParticleDataGroup:2022pth}
{\bfseries Particle Data Group} Collaboration, {\em {Review of Particle
  Physics},} \href{https://dx.doi.org/10.1093/ptep/ptac097}{PTEP {\bfseries
  2022} (2022) 083C01}.

\bibitem{Maksymyk:1993zm}
I.~Maksymyk, C.~P.~Burgess, and D.~London, {\em {Beyond S, T and U},}
  \href{https://dx.doi.org/10.1103/PhysRevD.50.529}{Phys.\  Rev.\  D {\bfseries
  50} (1994) 529--535} {\ttfamily
  [\href{https://arxiv.org/abs/hep-ph/9306267}{hep-ph/9306267}]}.

\bibitem{Barbieri:2004qk}
R.~Barbieri, A.~Pomarol, R.~Rattazzi, and A.~Strumia, {\em {Electroweak
  symmetry breaking after LEP-1 and LEP-2},}
  \href{https://dx.doi.org/10.1016/j.nuclphysb.2004.10.014}{Nucl.\  Phys.\  B
  {\bfseries 703} (2004) 127--146} {\ttfamily
  [\href{https://arxiv.org/abs/hep-ph/0405040}{hep-ph/0405040}]}.

\bibitem{Jaeckel:2010ni}
J.~Jaeckel and A.~Ringwald, {\em {The Low-Energy Frontier of Particle
  Physics},} \href{https://dx.doi.org/10.1146/annurev.nucl.012809.104433}{Ann.\
   Rev.\  Nucl.\  Part.\  Sci.\  {\bfseries 60} (2010) 405--437} {\ttfamily
  [\href{https://arxiv.org/abs/1002.0329}{arXiv:1002.0329}]}.

\bibitem{Cadamuro:2011fd}
D.~Cadamuro and J.~Redondo, {\em {Cosmological bounds on pseudo Nambu-Goldstone
  bosons},} \href{https://dx.doi.org/10.1088/1475-7516/2012/02/032}{JCAP
  {\bfseries 02} (2012) 032} {\ttfamily
  [\href{https://arxiv.org/abs/1110.2895}{arXiv:1110.2895}]}.

\bibitem{Proceedings:2012ulb}
J.~L.~Hewett {\em et~al.}, {\em {Fundamental Physics at the Intensity
  Frontier}.} {\ttfamily
  \href{https://arxiv.org/abs/1205.2671}{arXiv:1205.2671}}.

\bibitem{Izaguirre:2016dfi}
E.~Izaguirre, T.~Lin, and B.~Shuve, {\em {Searching for Axionlike Particles in
  Flavor-Changing Neutral Current Processes},}
  \href{https://dx.doi.org/10.1103/PhysRevLett.118.111802}{Phys.\  Rev.\
  Lett.\  {\bfseries 118} (2017) 111802} {\ttfamily
  [\href{https://arxiv.org/abs/1611.09355}{arXiv:1611.09355}]}.

\bibitem{Alonso-Alvarez:2018irt}
G.~Alonso-\'Alvarez, M.~B.~Gavela, and P.~Quilez, {\em {Axion couplings to
  electroweak gauge bosons},}
  \href{https://dx.doi.org/10.1140/epjc/s10052-019-6732-5}{Eur.\  Phys.\  J.\
  C {\bfseries 79} (2019) 223} {\ttfamily
  [\href{https://arxiv.org/abs/1811.05466}{arXiv:1811.05466}]}.

\bibitem{Gavela:2019wzg}
M.~B.~Gavela, R.~Houtz, P.~Quilez, R.~Del~Rey, and O.~Sumensari, {\em {Flavor
  constraints on electroweak ALP couplings},}
  \href{https://dx.doi.org/10.1140/epjc/s10052-019-6889-y}{Eur.\  Phys.\  J.\
  C {\bfseries 79} (2019) 369} {\ttfamily
  [\href{https://arxiv.org/abs/1901.02031}{arXiv:1901.02031}]}.

\bibitem{Guerrera:2021yss}
A.~W.~M.~Guerrera and S.~Rigolin, {\em {Revisiting $K \rightarrow \pi a$
  decays},} \href{https://dx.doi.org/10.1140/epjc/s10052-022-10146-x}{Eur.\
  Phys.\  J.\  C {\bfseries 82} (2022) 192} {\ttfamily
  [\href{https://arxiv.org/abs/2106.05910}{arXiv:2106.05910}]}.

\bibitem{Bauer:2021mvw}
M.~Bauer, M.~Neubert, S.~Renner, M.~Schnubel, and A.~Thamm, {\em {Flavor probes
  of axion-like particles},}
  \href{https://dx.doi.org/10.1007/JHEP09(2022)056}{JHEP {\bfseries 09} (2022)
  056} {\ttfamily [\href{https://arxiv.org/abs/2110.10698}{arXiv:2110.10698}]}.

\bibitem{Guerrera:2022ykl}
A.~W.~M.~Guerrera and S.~Rigolin, {\em {ALP production in weak mesonic
  decays},} \href{https://dx.doi.org/10.1002/prop.202200192}{Fortsch.\  Phys.\
  {\bfseries 2023} (2022) 2200192} {\ttfamily
  [\href{https://arxiv.org/abs/2211.08343}{arXiv:2211.08343}]}.

\bibitem{Mimasu:2014nea}
K.~Mimasu and V.~Sanz, {\em {ALPs at Colliders},}
  \href{https://dx.doi.org/10.1007/JHEP06(2015)173}{JHEP {\bfseries 06} (2015)
  173} {\ttfamily [\href{https://arxiv.org/abs/1409.4792}{arXiv:1409.4792}]}.

\bibitem{Jaeckel:2015jla}
J.~Jaeckel and M.~Spannowsky, {\em {Probing MeV to 90 GeV axion-like particles
  with LEP and LHC},}
  \href{https://dx.doi.org/10.1016/j.physletb.2015.12.037}{Phys.\  Lett.\  B
  {\bfseries 753} (2016) 482--487} {\ttfamily
  [\href{https://arxiv.org/abs/1509.00476}{arXiv:1509.00476}]}.

\bibitem{Jaeckel:2012yz}
J.~Jaeckel, M.~Jankowiak, and M.~Spannowsky, {\em {LHC probes the hidden
  sector},} \href{https://dx.doi.org/10.1016/j.dark.2013.06.001}{Phys.\  Dark
  Univ.\  {\bfseries 2} (2013) 111--117} {\ttfamily
  [\href{https://arxiv.org/abs/1212.3620}{arXiv:1212.3620}]}.

\bibitem{Bauer:2018uxu}
M.~Bauer, M.~Heiles, M.~Neubert, and A.~Thamm, {\em {Axion-Like Particles at
  Future Colliders},}
  \href{https://dx.doi.org/10.1140/epjc/s10052-019-6587-9}{Eur.\  Phys.\  J.\
  C {\bfseries 79} (2019) 74} {\ttfamily
  [\href{https://arxiv.org/abs/1808.10323}{arXiv:1808.10323}]}.

\bibitem{Florez:2021zoo}
A.~Fl\'orez, {\em et al.}, {\em {Probing axionlike particles with
  $\gamma\gamma$ final states from vector boson fusion processes at the LHC},}
  \href{https://dx.doi.org/10.1103/PhysRevD.103.095001}{Phys.\  Rev.\  D
  {\bfseries 103} (2021) 095001} {\ttfamily
  [\href{https://arxiv.org/abs/2101.11119}{arXiv:2101.11119}]}.

\bibitem{Wang:2021uyb}
D.~Wang, L.~Wu, J.~M.~Yang, and M.~Zhang, {\em {Photon-jet events as a probe of
  axionlike particles at the LHC},}
  \href{https://dx.doi.org/10.1103/PhysRevD.104.095016}{Phys.\  Rev.\  D
  {\bfseries 104} (2021) 095016} {\ttfamily
  [\href{https://arxiv.org/abs/2102.01532}{arXiv:2102.01532}]}.

\bibitem{dEnterria:2021ljz}
D.~d'Enterria in {\em {Workshop on Feebly Interacting Particles}}.
\newblock 2021.
\newblock {\ttfamily
  \href{https://arxiv.org/abs/2102.08971}{arXiv:2102.08971}}.

\bibitem{Knapen:2016moh}
S.~Knapen, T.~Lin, H.~K.~Lou, and T.~Melia, {\em {Searching for Axionlike
  Particles with Ultraperipheral Heavy-Ion Collisions},}
  \href{https://dx.doi.org/10.1103/PhysRevLett.118.171801}{Phys.\  Rev.\
  Lett.\  {\bfseries 118} (2017) 171801} {\ttfamily
  [\href{https://arxiv.org/abs/1607.06083}{arXiv:1607.06083}]}.

\bibitem{CMS:2018erd}
{\bfseries CMS} Collaboration, {\em {Evidence for light-by-light scattering and
  searches for axion-like particles in ultraperipheral PbPb collisions at
  $\sqrt{s_\mathrm{NN}} =$ 5.02 TeV},}
  \href{https://dx.doi.org/10.1016/j.physletb.2019.134826}{Phys.\  Lett.\  B
  {\bfseries 797} (2019) 134826} {\ttfamily
  [\href{https://arxiv.org/abs/1810.04602}{arXiv:1810.04602}]}.

\bibitem{ATLAS:2020hii}
{\bfseries ATLAS} Collaboration, {\em {Measurement of light-by-light scattering
  and search for axion-like particles with 2.2 nb$^{-1}$ of Pb+Pb data with the
  ATLAS detector},} \href{https://dx.doi.org/10.1007/JHEP11(2021)050}{JHEP
  {\bfseries 03} (2021) 243} {\ttfamily
  [\href{https://arxiv.org/abs/2008.05355}{arXiv:2008.05355}]}. [Erratum: JHEP
  11, 050 (2021)].

\bibitem{Craig:2018kne}
N.~Craig, A.~Hook, and S.~Kasko, {\em {The Photophobic ALP},}
  \href{https://dx.doi.org/10.1007/JHEP09(2018)028}{JHEP {\bfseries 09} (2018)
  028} {\ttfamily [\href{https://arxiv.org/abs/1805.06538}{arXiv:1805.06538}]}.

\bibitem{Bonilla:2022pxu}
J.~Bonilla, I.~Brivio, J.~Machado-Rodr\'\i{}guez, and J.~F.~de~Troc\'oniz, {\em
  {Nonresonant searches for axion-like particles in vector boson scattering
  processes at the LHC},}
  \href{https://dx.doi.org/10.1007/JHEP06(2022)113}{JHEP {\bfseries 06} (2022)
  113} {\ttfamily [\href{https://arxiv.org/abs/2202.03450}{arXiv:2202.03450}]}.

\bibitem{Passarino:1978jh}
G.~Passarino and M.~J.~G.~Veltman, {\em {One Loop Corrections for e+ e-
  Annihilation Into mu+ mu- in the Weinberg Model},}
  \href{https://dx.doi.org/10.1016/0550-3213(79)90234-7}{Nucl.\  Phys.\  B
  {\bfseries 160} (1979) 151--207}.

\bibitem{Georgi:1986df}
H.~Georgi, D.~B.~Kaplan, and L.~Randall, {\em {Manifesting the Invisible Axion
  at Low-energies},}
  \href{https://dx.doi.org/10.1016/0370-2693(86)90688-X}{Phys.\  Lett.\  B
  {\bfseries 169} (1986) 73--78}.

\bibitem{Bonilla:2021ufe}
J.~Bonilla, I.~Brivio, M.~B.~Gavela, and V.~Sanz, {\em {One-loop corrections to
  ALP couplings},} \href{https://dx.doi.org/10.1007/JHEP11(2021)168}{JHEP
  {\bfseries 11} (2021) 168} {\ttfamily
  [\href{https://arxiv.org/abs/2107.11392}{arXiv:2107.11392}]}.

\bibitem{Bauer:2020jbp}
M.~Bauer, M.~Neubert, S.~Renner, M.~Schnubel, and A.~Thamm, {\em {The
  Low-Energy Effective Theory of Axions and ALPs},}
  \href{https://dx.doi.org/10.1007/JHEP04(2021)063}{JHEP {\bfseries 04} (2021)
  063} {\ttfamily [\href{https://arxiv.org/abs/2012.12272}{arXiv:2012.12272}]}.

\bibitem{Arias-Aragon:2022iwl}
F.~Arias-Arag\'on, J.~Quevillon, and C.~Smith, {\em {Axion-like ALPs}.}
  {\ttfamily \href{https://arxiv.org/abs/2211.04489}{arXiv:2211.04489}}.

\bibitem{Aloni:2018vki}
D.~Aloni, Y.~Soreq, and M.~Williams, {\em {Coupling QCD-Scale Axionlike
  Particles to Gluons},}
  \href{https://dx.doi.org/10.1103/PhysRevLett.123.031803}{Phys.\  Rev.\
  Lett.\  {\bfseries 123} (2019) 031803} {\ttfamily
  [\href{https://arxiv.org/abs/1811.03474}{arXiv:1811.03474}]}.

\bibitem{Hagiwara:1994pw}
K.~Hagiwara, S.~Matsumoto, D.~Haidt, and C.~S.~Kim, {\em {A Novel approach to
  confront electroweak data and theory},}
  \href{https://dx.doi.org/10.1007/BF01957770}{Z.\  Phys.\  C {\bfseries 64}
  (1994) 559--620} {\ttfamily
  [\href{https://arxiv.org/abs/hep-ph/9409380}{hep-ph/9409380}]}. [Erratum:
  Z.Phys.C 68, 352 (1995)].

\bibitem{Hollik:1993cg}
W.~Hollik, {\em {Renormalization of the Standard Model},}
  \href{https://dx.doi.org/10.1142/9789814503662_0003}{Adv.\  Ser.\  Direct.\
  High Energy Phys.\  {\bfseries 14} (1995) 37--116}.

\bibitem{Hollik:1995dv}
W.~Hollik in {\em {5th Hellenic School and Workshops on Elementary Particle
  Physics}}.
\newblock 1995.
\newblock {\ttfamily
  \href{https://arxiv.org/abs/hep-ph/9602380}{hep-ph/9602380}}.

\bibitem{Cirigliano:2013lpa}
V.~Cirigliano and M.~J.~Ramsey-Musolf, {\em {Low Energy Probes of Physics
  Beyond the Standard Model},}
  \href{https://dx.doi.org/10.1016/j.ppnp.2013.03.002}{Prog.\  Part.\  Nucl.\
  Phys.\  {\bfseries 71} (2013) 2--20} {\ttfamily
  [\href{https://arxiv.org/abs/1304.0017}{arXiv:1304.0017}]}.

\bibitem{Cornwall:1981zr}
J.~M.~Cornwall, {\em {Dynamical Mass Generation in Continuum QCD},}
  \href{https://dx.doi.org/10.1103/PhysRevD.26.1453}{Phys.\  Rev.\  D
  {\bfseries 26} (1982) 1453}.

\bibitem{Cornwall:1989gv}
J.~M.~Cornwall and J.~Papavassiliou, {\em {Gauge Invariant Three Gluon Vertex
  in QCD},} \href{https://dx.doi.org/10.1103/PhysRevD.40.3474}{Phys.\  Rev.\  D
  {\bfseries 40} (1989) 3474}.

\bibitem{Degrassi:1992ue}
G.~Degrassi and A.~Sirlin, {\em {Gauge invariant selfenergies and vertex parts
  of the Standard Model in the pinch technique framework},}
  \href{https://dx.doi.org/10.1103/PhysRevD.46.3104}{Phys.\  Rev.\  D
  {\bfseries 46} (1992) 3104--3116}.

\bibitem{Degrassi:1992ff}
G.~Degrassi and A.~Sirlin, {\em {Gauge dependence of basic electroweak
  corrections of the standard model},}
  \href{https://dx.doi.org/10.1016/0550-3213(92)90671-W}{Nucl.\  Phys.\  B
  {\bfseries 383} (1992) 73--92}.

\bibitem{Binosi:2009qm}
D.~Binosi and J.~Papavassiliou, {\em {Pinch Technique: Theory and
  Applications},}
  \href{https://dx.doi.org/10.1016/j.physrep.2009.05.001}{Phys.\  Rept.\
  {\bfseries 479} (2009) 1--152} {\ttfamily
  [\href{https://arxiv.org/abs/0909.2536}{arXiv:0909.2536}]}.

\bibitem{deBlas:2022hdk}
J.~de~Blas, M.~Pierini, L.~Reina, and L.~Silvestrini, {\em {Impact of the
  recent measurements of the top-quark and W-boson masses on electroweak
  precision fits}.} {\ttfamily
  \href{https://arxiv.org/abs/2204.04204}{arXiv:2204.04204}}.

\bibitem{Janot:2019oyi}
P.~Janot and S.~Jadach, {\em {Improved Bhabha cross section at LEP and the
  number of light neutrino species},}
  \href{https://dx.doi.org/10.1016/j.physletb.2020.135319}{Phys.\  Lett.\
  {\bfseries B803} (2020) 135319}
{\ttfamily [\href{https://arxiv.org/abs/1912.02067}{arXiv:1912.02067}]}.

\bibitem{ALEPH:2005ab}
{\bfseries ALEPH, DELPHI, L3, OPAL, SLD, LEP Electroweak Working Group, SLD
  Electroweak Group, SLD Heavy Flavour Group} Collaboration, {\em {Precision
  electroweak measurements on the $Z$ resonance},}
  \href{https://dx.doi.org/10.1016/j.physrep.2005.12.006}{Phys.\ \ Rept.\
  {\bfseries 427} (2006) 257--454} {\ttfamily
  [\href{https://arxiv.org/abs/hep-ex/0509008}{hep-ex/0509008}]}.

\bibitem{Bernreuther:2016ccf}
W.~Bernreuther, L.~Chen, O.~Dekkers, T.~Gehrmann, and D.~Heisler, {\em {The
  forward-backward asymmetry for massive bottom quarks at the $Z$ peak at
  next-to-next-to-leading order QCD},}
  \href{https://dx.doi.org/10.1007/JHEP01(2017)053}{JHEP {\bfseries 01} (2017)
  053} {\ttfamily [\href{https://arxiv.org/abs/1611.07942}{arXiv:1611.07942}]}.

\bibitem{Schael:2013ita}
{\bfseries ALEPH, DELPHI, L3, OPAL, LEP Electroweak} Collaboration, {\em
  {Electroweak Measurements in Electron-Positron Collisions at W-Boson-Pair
  Energies at LEP},}
  \href{https://dx.doi.org/10.1016/j.physrep.2013.07.004}{Phys.\ \ Rept.\
  {\bfseries 532} (2013) 119--244} {\ttfamily
  [\href{https://arxiv.org/abs/1302.3415}{arXiv:1302.3415}]}.

\bibitem{Awramik:2003rn}
M.~Awramik, M.~Czakon, A.~Freitas, and G.~Weiglein, {\em {Precise prediction
  for the W boson mass in the standard model},}
  \href{https://dx.doi.org/10.1103/PhysRevD.69.053006}{Phys.\ \ Rev.\ \ D
  {\bfseries 69} (2004) 053006} {\ttfamily
  [\href{https://arxiv.org/abs/hep-ph/0311148}{hep-ph/0311148}]}.

\bibitem{Awramik:2006uz}
M.~Awramik, M.~Czakon, and A.~Freitas, {\em {Electroweak two-loop corrections
  to the effective weak mixing angle},}
  \href{https://dx.doi.org/10.1088/1126-6708/2006/11/048}{JHEP {\bfseries 11}
  (2006) 048} {\ttfamily
  [\href{https://arxiv.org/abs/hep-ph/0608099}{hep-ph/0608099}]}.

\bibitem{Dubovyk:2019szj}
I.~Dubovyk, A.~Freitas, J.~Gluza, T.~Riemann, and J.~Usovitsch, {\em
  {Electroweak pseudo-observables and Z-boson form factors at two-loop
  accuracy},} \href{https://dx.doi.org/10.1007/JHEP08(2019)113}{JHEP {\bfseries
  08} (2019) 113} {\ttfamily
  [\href{https://arxiv.org/abs/1906.08815}{arXiv:1906.08815}]}.

\bibitem{dEnterria:2020cpv}
D.~d'Enterria and V.~Jacobsen, {\em {Improved strong coupling determinations
  from hadronic decays of electroweak bosons at N$^3$LO accuracy}.} {\ttfamily
  \href{https://arxiv.org/abs/2005.04545}{arXiv:2005.04545}}.

\bibitem{FlavourLatticeAveragingGroupFLAG:2021npn}
{\bfseries Flavour Lattice Averaging Group (FLAG)} Collaboration, {\em {FLAG
  Review 2021},}
  \href{https://dx.doi.org/10.1140/epjc/s10052-022-10536-1}{Eur.\  Phys.\  J.\
  C {\bfseries 82} (2022) 869} {\ttfamily
  [\href{https://arxiv.org/abs/2111.09849}{arXiv:2111.09849}]}.

\bibitem{BaBar:2021ich}
{\bfseries BaBar} Collaboration, {\em {Search for an Axionlike Particle in $B$
  Meson Decays},}
  \href{https://dx.doi.org/10.1103/PhysRevLett.128.131802}{Phys.\  Rev.\
  Lett.\  {\bfseries 128} (2022) 131802} {\ttfamily
  [\href{https://arxiv.org/abs/2111.01800}{arXiv:2111.01800}]}.

\bibitem{LHCb:2016awg}
{\bfseries LHCb} Collaboration, {\em {Search for long-lived scalar particles in
  $B^+ \to K^+ \chi (\mu^+\mu^-)$ decays},}
  \href{https://dx.doi.org/10.1103/PhysRevD.95.071101}{Phys.\  Rev.\  D
  {\bfseries 95} (2017) 071101} {\ttfamily
  [\href{https://arxiv.org/abs/1612.07818}{arXiv:1612.07818}]}.

\bibitem{Belle-II:2020jti}
{\bfseries Belle-II} Collaboration, {\em {Search for Axion-Like Particles
  produced in $e^+e^-$ collisions at Belle II},}
  \href{https://dx.doi.org/10.1103/PhysRevLett.125.161806}{Phys.\  Rev.\
  Lett.\  {\bfseries 125} (2020) 161806} {\ttfamily
  [\href{https://arxiv.org/abs/2007.13071}{arXiv:2007.13071}]}.

\bibitem{CDF:2013lma}
{\bfseries CDF} Collaboration, {\em {First Search for Exotic Z Boson Decays
  into Photons and Neutral Pions in Hadron Collisions},}
  \href{https://dx.doi.org/10.1103/PhysRevLett.112.111803}{Phys.\  Rev.\
  Lett.\  {\bfseries 112} (2014) 111803} {\ttfamily
  [\href{https://arxiv.org/abs/1311.3282}{arXiv:1311.3282}]}.

\bibitem{ATLAS:2015rsn}
{\bfseries ATLAS} Collaboration, {\em {Search for new phenomena in events with
  at least three photons collected in $pp$ collisions at $\sqrt{s}$ = 8 TeV
  with the ATLAS detector},}
  \href{https://dx.doi.org/10.1140/epjc/s10052-016-4034-8}{Eur.\  Phys.\  J.\
  C {\bfseries 76} (2016) 210} {\ttfamily
  [\href{https://arxiv.org/abs/1509.05051}{arXiv:1509.05051}]}.

\bibitem{ATLAS:2017ayi}
{\bfseries ATLAS} Collaboration, {\em {Search for new phenomena in high-mass
  diphoton final states using 37 fb$^{-1}$ of proton--proton collisions
  collected at $\sqrt{s}=13$ TeV with the ATLAS detector},}
  \href{https://dx.doi.org/10.1016/j.physletb.2017.10.039}{Phys.\  Lett.\  B
  {\bfseries 775} (2017) 105--125} {\ttfamily
  [\href{https://arxiv.org/abs/1707.04147}{arXiv:1707.04147}]}.

\end{thebibliography}\endgroup
\end{document}